\newcommand{\CLASSINPUTtoptextmargin}{2.28cm}
\newcommand{\CLASSINPUTbottomtextmargin}{2.45cm}
\documentclass[conference]{IEEEtran}

\usepackage{xcolor}
\definecolor{links}{rgb}{0.5,0,0}   
\definecolor{urls}{rgb}{0,0,0.8}    
\definecolor{cites}{rgb}{0,0,0.6}   
\usepackage[colorlinks,hyperindex,linkcolor=links,citecolor=cites,urlcolor=urls]{hyperref} 

\usepackage{graphicx}
\usepackage{amsmath}
\usepackage{balance}
\usepackage{cite}
\usepackage{url}
\usepackage{bbm}

\usepackage{vmr-symbols-vecbold}
\usepackage{standard-macros}
\PassOptionsToPackage{hyphens}{url}
\usepackage{hyperref}

\DeclareSymbolFontAlphabet{\amsmathbb}{AMSb}%

\newcommand{\lro}[1]{\lefto({#1}\right)}																
\newcommand{\lrbo}[1]{\lefto \lbrace {#1} \right \rbrace}															
\newcommand{\lrho}[1]{\lefto [ {#1} \right ]}																				

\newcommand{\lr}[1]{\left({#1}\right)}																
\newcommand{\lrb}[1]{\left \lbrace {#1} \right \rbrace}															

\safemath{\dopplerspread}{B_D}																								
\safemath{\delayspread}{T_D}																									
\safemath{\nc}{n\sub{c}}																										
\safemath{\nf}{n\sub{f}}																										
\safemath{\efa}{p\sub{sc}}
\safemath{\efb}{p\sub{cs}}
\safemath{\ef}{\epsilon\sub{f}	}
\safemath{\nd}{n\sub{d}}																										
\safemath{\ntx}{n\sub{t}} 																											
\safemath{\nrx}{n\sub{r}}																											
\safemath{\ntxt}{\tilde{n\sub{t}}}																											
\safemath{\cb}{\ensuremath{L}} 																								
\safemath{\cl}{\ensuremath{n}} 																								
\safemath{\txanto}{{\ensuremath{\tilde{m}_t}}} 																		
\safemath{\cs}{M} 																														
\safemath{\idPustm}{\ensuremath{S_{k}}}
\safemath{\error}{\ensuremath{\epsilon}} 																				
\safemath{\eexp}{\ensuremath{\mathcal{E}}} 																			
\safemath{\nsubc}{n\sub{s}}			 																						
\safemath{\nofdm}{n\sub{o}} 																									
\safemath{\bc}{\ensuremath{B_c}} 																							
\safemath{\ts}{\ensuremath{T_s}} 																							
\safemath{\nrb}{\ensuremath{n_{rb}}} 																						
\safemath{\rul}{\ensuremath{\rho\sub{ul}}}
\safemath{\rdl}{\ensuremath{\rho\sub{dl}}}

\safemath{\nres}{\ell}
\safemath{\nr}{n\sub{r}}
\safemath{\maxk}{M^*\lr{\nres, \nsubc, \nofdm, \epsilon, \rho}}
\safemath{\Rmax}{R^*}
\safemath{\Emin}{E\sub{b}^*/N_0}
\safemath{\Eminf}{\frac{E\sub{b}^*}{N_0}}
\safemath{\np}{\ensuremath{n\sub{p}}}
\safemath{\ndf}{\ensuremath{\bar{n}\sub{d}}}
\safemath{\npf}{\ensuremath{\bar{n}\sub{p}}}
\safemath{\code}{\ensuremath{\mathcal{C}}}
\safemath{\err}{\ensuremath{\epsilon}}
\safemath{\rp}{\ensuremath{\rho\sub{p}}}
\safemath{\rd}{\ensuremath{\rho\sub{d}}}
\safemath{\cohtime}{\ensuremath{T\sub{c}}}
\safemath{\cohbw}{\ensuremath{B\sub{c}}}
\safemath{\nmax}{\ensuremath{\ell\sub{m}}}
\safemath{\ntot}{\ensuremath{n\sub{tot}}}
\safemath{\nul}{\ensuremath{n\sub{ul}}}
\safemath{\ndl}{\ensuremath{n\sub{dl}}}

\safemath{\yp}{\ensuremath{\randvecy_{\nu}^{(\text{p})}}}
\safemath{\yd}{\ensuremath{\randvecy_{\nu}^{(\text{d})}}}
\safemath{\ypd}{\ensuremath{\vecy_{\nu}^{(\text{p})}}}
\safemath{\ydd}{\ensuremath{\vecy_{\nu}^{(\text{d})}}}

\safemath{\ypf}{\ensuremath{\bar{\randvecy}_{\nu}^{(\text{p})}}}
\safemath{\ydf}{\ensuremath{\bar{\randvecy}_{\nu}^{(\text{d})}}}
\safemath{\ypdf}{\ensuremath{\bar{\vecy}_{\nu}^{(\text{p})}}}
\safemath{\yddf}{\ensuremath{\bar{\vecy}_{\nu}^{(\text{d})}}}

\safemath{\xp}{\ensuremath{\vecx^{(\text{p})}}}
\safemath{\xd}{\ensuremath{\randvecx_{\nu}^{(\text{d})}}}
\safemath{\xdd}{\ensuremath{\vecx_{\nu}^{(\text{d})}}}

\safemath{\xpf}{\ensuremath{\bar{\vecx}^{(\text{p})}}}
\safemath{\xdf}{\ensuremath{\bar{\randvecx}_{\nu}^{(\text{d})}}}
\safemath{\xddf}{\ensuremath{\bar{\vecx}_{\nu}^{(\text{d})}}}

\safemath{\xdb}{\ensuremath{\overline{\randvecx}^{(\text{d})}}}
\safemath{\Pxd}{\ensuremath{P_{\randvecx^{(\text{d})}}}}

\safemath{\xpbar}{\ensuremath{\overline{\matX}^{(\text{p})}}}
\safemath{\xdbar}{\ensuremath{\overline{\randmatX}^{(\text{d})}}}

\safemath{\xdv}{\ensuremath{\randvecx^{(\text{d})}}}
\safemath{\xdbarv}{\ensuremath{\overline{\randvecx}^{(\text{d})}}}
\safemath{\ydv}{\ensuremath{\randvecy^{(\text{d})}}}

\safemath{\xdr}{\ensuremath{\matX^{(\text{d})}}}

\safemath{\ttx}{\ensuremath{\tau\sub{tx}}}
\safemath{\trx}{\ensuremath{\tau\sub{rx}}}
\safemath{\ack}{\ensuremath{\mathrm{s}}}
\safemath{\nack}{\ensuremath{\mathrm{c}}}

\newcommand{\prob}[1]{\ensuremath{\mathbb{P}\lrho{#1}}}

\safemath{\mI}{\ensuremath{i\lro{\randvecy ; \randvecx}}} 				

\safemath{\randveca}{\bm{A}}
\safemath{\randvecb}{\bm{B}}
\safemath{\randvecc}{\bm{C}}
\safemath{\randvecd}{\bm{D}}
\safemath{\randvece}{\bm{E}}
\safemath{\randvecf}{\bm{F}}
\safemath{\randvecg}{\bm{G}}
\safemath{\randvech}{\bm{H}}
\safemath{\randveci}{\bm{I}}
\safemath{\randvecj}{\bm{J}}
\safemath{\randveck}{\bm{K}}
\safemath{\randvecl}{\bm{L}}
\safemath{\randvecm}{\bm{M}}
\safemath{\randvecn}{\bm{N}}
\safemath{\randveco}{\bm{O}}
\safemath{\randvecp}{\bm{P}}
\safemath{\randvecq}{\bm{Q}}
\safemath{\randvecr}{\bm{R}}
\safemath{\randvecs}{\bm{S}}
\safemath{\randvect}{\bm{T}}
\safemath{\randvecu}{\bm{U}}
\safemath{\randvecv}{\bm{V}}
\safemath{\randvecw}{\bm{W}}
\safemath{\randvecx}{\bm{X}}
\safemath{\randvecy}{\bm{Y}}
\safemath{\randvecz}{\bm{Z}}
\safemath{\randvecphi}{\bm{\Phi}}

\safemath{\randmatA}{\amsmathbb{A}}
\safemath{\randmatB}{\amsmathbb{B}}
\safemath{\randmatC}{\amsmathbb{C}}
\safemath{\randmatD}{\amsmathbb{D}}
\safemath{\randmatE}{\amsmathbb{E}}
\safemath{\randmatF}{\amsmathbb{F}}
\safemath{\randmatG}{\amsmathbb{G}}
\safemath{\randmatH}{\amsmathbb{H}}
\safemath{\randmatI}{\amsmathbb{I}}
\safemath{\randmatJ}{\amsmathbb{J}}
\safemath{\randmatK}{\amsmathbb{K}}
\safemath{\randmatL}{\amsmathbb{L}}
\safemath{\randmatM}{\amsmathbb{M}}
\safemath{\randmatN}{\amsmathbb{N}}
\safemath{\randmatO}{\amsmathbb{O}}
\safemath{\randmatP}{\amsmathbb{P}}
\safemath{\randmatQ}{\amsmathbb{Q}}
\safemath{\randmatR}{\amsmathbb{R}}
\safemath{\randmatS}{\amsmathbb{S}}
\safemath{\randmatT}{\amsmathbb{T}}
\safemath{\randmatU}{\amsmathbb{U}}
\safemath{\randmatV}{\amsmathbb{V}}
\safemath{\randmatW}{\amsmathbb{W}}
\safemath{\randmatX}{\amsmathbb{X}}
\safemath{\randmatY}{\amsmathbb{Y}}
\safemath{\randmatZ}{\amsmathbb{Z}}
\safemath{\randmatSigma}{\mathbb{\Sigma}}
\safemath{\randmatPhi}{\mathbb{\Phi}}
\safemath{\randmatLambda}{\mathbb{\Lambda}}

\safemath{\matSigma}{\bm{\Sigma}}
\safemath{\matPhi}{\bm{\Phi}}
\safemath{\matLambda}{\bm{\Lambda}}

\long\def\comment#1{}

\newcommand{\beq}{\begin{equation}}
\newcommand{\eeq}{\end{equation}}

\newcommand{\mv}{{\bf m}}

\newcommand{\Lc}{{\cal L}}

\renewcommand{\det}{{\hbox{det}}}

\usepackage{glossaries}
\glsdisablehyper
\newglossaryentry{simo}
{
  name={SIMO},
  description={single-input multiple-output},
  first={\glsentrydesc{simo} (\glsentrytext{simo})}
}

\newglossaryentry{mmse}
{
  name={MMSE},
  description={minimum mean-square error},
  first={\glsentrydesc{mmse} (\glsentrytext{mmse})}
}

\newglossaryentry{na}
{
  name={NA},
  description={normal approximation},
  first={\glsentrydesc{na} (\glsentrytext{na})},
  plural={NAs},
  descriptionplural={normal approximations},
  firstplural={\glsentrydescplural{na} (\glsentryplural{na})}
}

\newglossaryentry{zf}
{
  name={ZF},
  description={zero-forcing},
  first={\glsentrydesc{zf} (\glsentrytext{zf})}
}

\newglossaryentry{rzf}
{
  name={RZF},
  description={regularized zero-forcing},
  first={\glsentrydesc{rzf} (\glsentrytext{rzf})}
}

\newglossaryentry{sir}
{
  name={SIR},
  description={signal-to-interference ratio},
  first={\glsentrydesc{sir} (\glsentrytext{sir})}
}

\newglossaryentry{mse}
{
  name={MSE},
  description={mean-square error},
  first={\glsentrydesc{mse} (\glsentrytext{mse})}
}
\newglossaryentry{mmmse}
{
  name={M-MMSE},
  description={multicell MMSE},
  first={\glsentrydesc{mmmse} (\glsentrytext{mmmse})}
}

\newglossaryentry{ls}
{
  name={LS},
  description={least-squares},
  first={\glsentrydesc{ls} (\glsentrytext{ls})}
}

\newglossaryentry{mr}
{
  name={MR},
  description={maximum ratio},
  first={\glsentrydesc{mr} (\glsentrytext{mr})}
}

\newglossaryentry{ue}
{
  name={UE},
  description={user equipment},
  first={\glsentrydesc{ue} (\glsentrytext{ue})},
  plural={UEs},
  descriptionplural={user equipments},
  firstplural={\glsentrydescplural{ue} (\glsentryplural{ue})}
}

\newglossaryentry{ap}
{
  name={AP},
  description={access point},
  first={\glsentrydesc{ap} (\glsentrytext{ap})},
  plural={APs},
  descriptionplural={access points},
  firstplural={\glsentrydescplural{ap} (\glsentryplural{ap})}
}

\newglossaryentry{cpu}
{
  name={CPU},
  description={central processing unit},
  first={\glsentrydesc{cpu} (\glsentrytext{cpu})},
  plural={CPUs},
  descriptionplural={central processing units},
  firstplural={\glsentrydescplural{cpu} (\glsentryplural{cpu})}
}

\newglossaryentry{bs}
{
  name={BS},
  description={base station},
  first={\glsentrydesc{bs} (\glsentrytext{bs})},
  plural={BSs},
  descriptionplural={base stations},
  firstplural={\glsentrydescplural{bs} (\glsentryplural{bs})}
}

\newglossaryentry{ostbc}
{
  name={OSTBC},
  description={orthogonal space-time block code},
  first={\glsentrydesc{ostbc} (\glsentrytext{ostbc})},
  plural={OSTBCs},
  descriptionplural={orthogonal space-time block codes},
  firstplural={\glsentrydescplural{ostbc} (\glsentryplural{ostbc})}
}

\newglossaryentry{biawgn}
{
  name={bi-AWGN},
  description={binary-input additive white Gaussian noise},
  first={\glsentrydesc{biawgn} (\glsentrytext{biawgn})}
}

\newglossaryentry{flnf}
{
  name={FLNF},
  description={fixed-length no-feedback},
  first={\glsentrydesc{flnf} (\glsentrytext{flnf})}
}

\newglossaryentry{crc}
{
  name={CRC},
  description={cyclic redundancy check},
  first={\glsentrydesc{crc} (\glsentrytext{crc})}
}

\newglossaryentry{bsc}
{
  name={BSC},
  description={binary symmetric channel},
  first={\glsentrydesc{bsc} (\glsentrytext{bsc})}
  plural={BSCs},
  descriptionplural={binary symmetric channels},
  firstplural={\glsentrydescplural{bsc} (\glsentryplural{bsc})}
}

\newglossaryentry{bec}
{
  name={BEC},
  description={binary-erasure channel},
  first={\glsentrydesc{bec} (\glsentrytext{bec})}
}
\newglossaryentry{awgn}
{
  name={AWGN},
  description={additive white Gaussian noise},
  first={\glsentrydesc{awgn} (\glsentrytext{awgn})}
}
\newglossaryentry{3gpp}
{
  name={3GPP},
  description={3rd generation partnership project},
  first={\glsentrydesc{3gpp} (\glsentrytext{3gpp})}
}

\newglossaryentry{dl}
{
  name={DL},
  description={downlink},
  first={\glsentrydesc{dl} (\glsentrytext{dl})}
}

\newglossaryentry{LED}
{
  name={LED},
  description={light emitting diode},
  first={\glsentrydesc{LED} (\glsentrytext{LED})},
  plural={LEDs},
  descriptionplural={light emitting diodes},
  firstplural={\glsentrydescplural{LED} (\glsentryplural{LED})}
}

\newglossaryentry{lte}
{
  name={LTE},
  description={long term evolution},
  first={\glsentrydesc{lte} (\glsentrytext{lte})}
}
\newglossaryentry{ofdm}
{
  name={OFDM},
  description={orthogonal frequency-division multiplexing},
  first={\glsentrydesc{ofdm} (\glsentrytext{ofdm})}
}

\newglossaryentry{ul}
{
  name={UL},
  description={uplink},
  first={\glsentrydesc{ul} (\glsentrytext{ul})}
}
\newglossaryentry{rb}
{
  name={RB},
  description={resource block},
  first={\glsentrydesc{rb} (\glsentrytext{rb})},
  plural={RBs},
  descriptionplural={resource blocks},
  firstplural={\glsentrydescplural{rb} (\glsentryplural{rb})}
}
\newglossaryentry{cu}
{
  name={CU},
  description={channel use},
  first={\glsentrydesc{cu} (\glsentrytext{cu})},
  plural={CUs},
  descriptionplural={channel uses},
  firstplural={\glsentrydescplural{cu} (\glsentryplural{cu})}
}
\newglossaryentry{tti}
{
  name={TTI},
  description={transmission time interval},
  first={\glsentrydesc{tti} (\glsentrytext{tti})},
  plural={TTI's},
  descriptionplural={transmission time intervals},
  firstplural={\glsentrydescplural{tti} (\glsentryplural{tti})}
}
\newglossaryentry{iid}
{
  name={i.i.d.},
  description={independent and identically distributed},
  first={\glsentrydesc{iid} (\glsentrytext{iid})}
}
\newglossaryentry{psd}
{
  name={PSD},
  description={power spectral density},
  first={\glsentrydesc{psd} (\glsentrytext{psd})}
}

\newglossaryentry{mimo}
{
  name={MIMO},
  description={multiple-input multiple-output},
  first={\glsentrydesc{mimo} (\glsentrytext{mimo})}
}

\newglossaryentry{bler}
{
  name={BLER},
  description={block error probability},
  first={\glsentrydesc{bler} (\glsentrytext{bler})}
}
\newglossaryentry{mtc}
{
  name={MTC},
  description={machine-type communication},
  first={\glsentrydesc{mtc} (\glsentrytext{mtc})}
}
\newacronym{csi}{CSI}{channel state information}
\newglossaryentry{dvbt}
{
  name={DVB-T},
  description={terrestrial digital video broadcasting},
  first={\glsentrydesc{dvbt} (\glsentrytext{dvbt})}
}
\newglossaryentry{pat}
{
  name={PAT},
  description={pilot-assisted transmission},
  first={\glsentrydesc{pat} (\glsentrytext{pat})}
}
\newglossaryentry{ml}
{
  name={ML},
  description={maximum likelihood},
  first={\glsentrydesc{ml} (\glsentrytext{ml})}
}
\newglossaryentry{dt}
{
  name={DT},
  description={dependence-testing},
  first={\glsentrydesc{dt} (\glsentrytext{dt})}
}
\newglossaryentry{ustm}
{
  name={USTM},
  description={unitary space-time modulation},
  first={\glsentrydesc{ustm} (\glsentrytext{ustm})}
}
\newglossaryentry{siso}
{
  name={SISO},
  description={single-input single-output},
  first={\glsentrydesc{siso} (\glsentrytext{siso})}
}
\newglossaryentry{snn}
{
  name={SNN},
  description={scaled nearest-neighbor},
  first={\glsentrydesc{snn} (\glsentrytext{snn})}
}

\newglossaryentry{snnd}
{
  name={SNND},
  description={scaled nearest-neighbor decoding},
  first={\glsentrydesc{snnd} (\glsentrytext{snnd})}
}
\newglossaryentry{rcus}
{
  name={RCUs},
  description={random-coding union bound with parameter $s$},
  first={\glsentrydesc{rcus} (\glsentrytext{rcus})}
}
\newglossaryentry{osd}
{
  name={OSD},
  description={ordered statistics decoding},
  first={\glsentrydesc{osd} (\glsentrytext{osd})}
}
\newglossaryentry{llr}
{
  name={LLR},
  description={log-likelihood ratio},
  first={\glsentrydesc{llr} (\glsentrytext{llr})}
   plural={LLR's},
  descriptionplural={log-likelihood ratios},
  firstplural={\glsentrydescplural{llr} (\glsentryplural{llr})}
}
\newglossaryentry{qpsk}
{
  name={QPSK},
  description={quaternary phase shift keying},
  first={\glsentrydesc{qpsk} (\glsentrytext{qpsk})}
}
\newglossaryentry{nlos}
{
  name={NLOS},
  description={non-line-of-sight},
  first={\glsentrydesc{nlos} (\glsentrytext{nlos})}
}
\newglossaryentry{los}
{
  name={LOS},
  description={line-of-sight},
  first={\glsentrydesc{los} (\glsentrytext{los})}
}
\newglossaryentry{mgf}
{
  name={MGF},
  description={moment-generating function},
  first={\glsentrydesc{mgf} (\glsentrytext{mgf})}
}
\newglossaryentry{cgf}
{
  name={CGF},
  description={cumulant-generating function},
  first={\glsentrydesc{cgf} (\glsentrytext{cgf})},
  plural={CGFs},
  descriptionplural={cumulant-generating functions},
  firstplural={\glsentrydescplural{cgf} (\glsentryplural{cgf})}
}
\newglossaryentry{pdf}
{
  name={PDF},
  description={probability density function},
  first={\glsentrydesc{pdf} (\glsentrytext{pdf})},
   plural={PDF's},
  descriptionplural={probability density functions},
  firstplural={\glsentrydescplural{pdf} (\glsentryplural{pdf})}
}

\newglossaryentry{pmf}
{
  name={PMF},
  description={probability mass function},
  first={\glsentrydesc{pmf} (\glsentrytext{pmf})},
   plural={PMF's},
  descriptionplural={probability mass functions},
  firstplural={\glsentrydescplural{pmf} (\glsentryplural{pmf})}
}

\newglossaryentry{ccdf}
{
  name={CCDF},
  description={complementary cumulative distribution function},
  first={\glsentrydesc{ccdf} (\glsentrytext{ccdf})}
}

\newglossaryentry{cdf}
{
  name={CDF},
  description={cumulative distribution function},
  first={\glsentrydesc{cdf} (\glsentrytext{cdf})}
}

\newglossaryentry{gmi}
{
  name={GMI},
  description={generalized mutual information},
  first={\glsentrydesc{gmi} (\glsentrytext{gmi})}
}

\newglossaryentry{arq}
{
  name={ARQ},
  description={automatic repeat request},
  first={\glsentrydesc{arq} (\glsentrytext{arq})}
}

\newglossaryentry{harq}
{
  name={HARQ},
  description={hybrid automatic repeat request},
  first={\glsentrydesc{harq} (\glsentrytext{harq})}
}

\newglossaryentry{fbl}
{
  name={FBL-NF},
  description={fixed blocklength with no feedback},
  first={\glsentrydesc{fbl} (\glsentrytext{fbl})}
}
\newglossaryentry{ssnf}
{
  name={SS-NF},
  description={single-shot no-feedback},
  first={\glsentrydesc{ssnf} (\glsentrytext{ssnf})}
}

\newacronym{urllc}{URLLC}{ultra-reliable low-latency communications}

\newglossaryentry{vlsf}
{
  name={VLSF},
  description={Variable-length stop-feedback},
  first={\glsentrydesc{vlsf} (\glsentrytext{vlsf})}
}

\newglossaryentry{vlnsf}
{
  name={VLNSF},
  description={variable-length noisy stop feedback},
  first={\glsentrydesc{vlnsf} (\glsentrytext{vlnsf})}
}

\newglossaryentry{dmt}
{
  name={DMT},
  description={diversity-multiplexing tradeoff},
  first={\glsentrydesc{dmt} (\glsentrytext{dmt})}
}

\newglossaryentry{dmdt}
{
  name={DMDT},
  description={diversity-multiplexing-delay tradeoff},
  first={\glsentrydesc{dmdt} (\glsentrytext{dmdt})}
}

\newglossaryentry{tddt}
{
  name={TDDT},
  description={throughput-diversity-delay tradeoff},
  first={\glsentrydesc{tddt} (\glsentrytext{tddt})}
}

\newglossaryentry{tdd}
{
  name={TDD},
  description={time-division duplex},
  first={\glsentrydesc{tdd} (\glsentrytext{tdd})}
}

\newglossaryentry{stbc}
{
  name={STBC},
  description={space-time block-codes},
  first={\glsentrydesc{stbc} (\glsentrytext{stbc})},
  plural={STBC's},
  descriptionplural={space-time block-codes},
  firstplural={\glsentrydescplural{stbc} (\glsentryplural{stbc})}
}

\newglossaryentry{harqir}
{
  name={HARQ-IR},
  description={hybrid automatic repeat request with incremental redundancy},
  first={\glsentrydesc{harqir} (\glsentrytext{harqir})}
}

\newglossaryentry{harqcc}
{
  name={HARQ-CC},
  description={hybrid automatic repeat request with chase combining},
  first={\glsentrydesc{harqcc} (\glsentrytext{harqcc})}
}

\newglossaryentry{wp1}
{
  name={w.p.1.},
  description={with probability one},
  first={\glsentrydesc{wp1} (\glsentrytext{wp1})}
}

\newglossaryentry{vlff}
{
  name={VLFF},
  description={variable-length full feedback},
  first={\glsentrydesc{vlff} (\glsentrytext{vlff})}
}

\newglossaryentry{dmc}
{
  name={DMC},
  description={discrete memoryless channel},
  first={\glsentrydesc{dmc} (\glsentrytext{dmc})}
  plural={DMCs},
  descriptionplural={discrete memoryless channels},
  firstplural={\glsentrydescplural{dmc} (\glsentryplural{dmc})}
}

\newglossaryentry{bac}
{
  name={BAC},
  description={binary-adder channel},
  first={\glsentrydesc{bac} (\glsentrytext{bac})}
}

\newglossaryentry{tin}
{
  name={TIN},
  description={treating interference as noise},
  first={\glsentrydesc{tin} (\glsentrytext{tin})}
}

\newglossaryentry{pupe}
{
  name={PUPE},
  description={per-user probability of error},
  first={\glsentrydesc{pupe} (\glsentrytext{pupe})}
}

\newglossaryentry{jpe}
{
  name={JPE},
  description={joint probability of error},
  first={\glsentrydesc{jpe} (\glsentrytext{jpe})}
}

\newglossaryentry{sparc}
{
  name={SPARC},
  description={sparse regression code},
  first={\glsentrydesc{sparc} (\glsentrytext{sparc})},
  plural={SPARCs},
  descriptionplural={sparse regression codes},
  firstplural={\glsentrydescplural{sparc} (\glsentryplural{sparc})}
}

\newglossaryentry{mac}
{
  name={MAC},
  description={multiple-access channel},
  first={\glsentrydesc{mac} (\glsentrytext{mac})},
  plural={MACs},
  descriptionplural={multiple access channels},
  firstplural={\glsentrydescplural{mac} (\glsentryplural{mac})}
}

\newglossaryentry{mmtc}
{
  name={mMTC},
  description={massive machine-type communications},
  first={\glsentrydesc{mmtc} (\glsentrytext{mmtc})},
  plural={mMTCs},
  descriptionplural={massive machine type communication},
  firstplural={\glsentrydescplural{mmtc} (\glsentryplural{mmtc})}
}

\newglossaryentry{ura}
{
  name={URA},
  description={unsourced random access},
  first={\glsentrydesc{ura} (\glsentrytext{ura})}
}

\newglossaryentry{ccs}
{
  name={CCS},
  description={coded compressed sensing},
  first={\glsentrydesc{ccs} (\glsentrytext{ccs})}
}
\usepackage{pgfplots}
\pgfplotsset{compat=1.14}
\usepgfplotslibrary{groupplots}
\usetikzlibrary{pgfplots.groupplots} 
\usepackage{subcaption}
\newtheorem{theorem}{Theorem}

\newtheorem{definition}{Definition}

\newtheorem{corollary}{Corollary}

\newtheorem{remark}{Remark}

\newcommand\eqdist{\stackrel{\mbox{\small$d$}}{=}}

\let\comment\undefined

\IEEEoverridecommandlockouts
\begin{document}

\title{Finite-Blocklength Results for the A-channel: Applications to Unsourced Random Access and Group Testing}
\author{Alejandro Lancho, Alexander Fengler and Yury Polyanskiy
\thanks{The authors are with the Department of Electrical Engineering and Computer Science, Massachusetts Institute of Technology, Cambridge 02139, MA, USA (e-mails: \{lancho,fengler,yp\}@mit.edu). 
        Alejandro Lancho has received funding from the European Union's Horizon 2020 research and innovation programme under the Marie Sklodowska-Curie grant agreement No. 101024432. Alexander Fengler was funded by the Deutsche Forschungsgemeinschaft (DFG, German Research Foundation) – Grant 471512611. This work is also supported by the National Science Foundation under Grant No CCF-2131115.
        }
 }
 \maketitle
 \sloppy 
\begin{abstract}
  We present finite-blocklength achievability bounds for the \emph{unsourced} A-channel. In this multiple-access channel, users noiselessly transmit codewords picked from a common codebook with entries generated from a $q$-ary alphabet. At each channel use, the receiver observes the set of different transmitted symbols but not their multiplicity. 
  We show that the A-channel finds applications in unsourced random-access (URA)  and group testing. Leveraging the insights provided by the finite-blocklength bounds and the connection between URA and non-adaptive group testing through the A-channel, we propose improved decoding methods for state-of-the-art A-channel codes and we showcase how A-channel codes provide a new class of structured group testing matrices. The developed bounds allow to evaluate the achievable error probabilities of group testing matrices based on random A-channel codes for arbitrary numbers of tests, items and defectives. We show that such a construction asymptotically achieves the optimal number of tests. In addition, every efficiently decodable A-channel code can be used to construct a group testing matrix with sub-linear recovery time.
\end{abstract}



%
\section{Introduction}\label{sec:intro}
We consider the problem where $K$ users transmit symbols from a $q$-ary input alphabet $[q] =\{1,\ldots,q\}$ over a noiseless channel. Specifically, let $c_{i,j} \in [q]$ be the transmitted symbol from user $j \in [K]$ at channel use $i$. The channel output $Y_i$ at channel use $i$ is given by
\begin{equation}\label{eq:channel1}
    Y_i = \bigcup_{j=1}^{K}c_{i,j}.
\end{equation} 
In this channel, sometimes referred to as A-channel~\cite{Chang81,Bassalygo2000}, the receiver observes the set of transmitted symbols but not who transmitted them, and also not the multiplicity.\footnote{Note that, for the case where $K=2$, the multiplicity can be inferred from the cardinality of $Y$, and thus, for $q=2$, the A-channel is equivalent to the \gls{bac}.} 
 The A-channel was introduced by Chang and Wolf in~\cite{Chang81} as the ``$T$-user $M$-frequency channel without intensity information'', and it is also known as the hyperchannel \cite{Bas2013}. The mutual information of the A-channel under uniform inputs was obtained in \cite{Chang81}. Its limit when $K$ and $q$ tend to infinity but its ratio $\lambda = K/q$ is fixed was studied in~\cite{Bassalygo2000}.
 Specifically, in~\cite{Bassalygo2000}, it was shown that in this limit the mutual information grows proportional to $q$.
 Also in~\cite{Bassalygo2000}, it was shown that uniform inputs are not optimal in general, although they become optimal in the limit $\lambda\to 0$ and when $\lambda = \ln 2$, where $\ln(\cdot)$ denotes the natural logarithm.
 Besides, when the input distributions of the users are constrained to be equal, uniform distributions become asymptotically optimal for all $\lambda \leq \ln 2$ \cite{Bassalygo2000}.
 The mutual information with uniform inputs in the sparse limit of $K,q\to \infty$ with fixed ratio $(\log K)/q$ was computed in
 \cite{Fengler21-05} and it was shown that in this limit the mutual information grows proportional to $\log q$. Furthermore, in this regime the simplified \emph{cover} decoder, which checks each codeword individually for consistency with the channel output, is optimal.
 For general $K$ and $q$, the optimal input distribution as well as the capacity of the A-channel are still unknown. 
 
In the case where all users transmit their messages from a common codebook, we will refer to \eqref{eq:channel1} as the \emph{unsourced} A-channel. Under this setup, 
the receiver can only recover a list of transmitted codewords up to permutation.
The information theoretic question of multiple-access in the unsourced setting was
first formulated in \cite{polyanskiy17-06a} for the AWGN \gls{mac}, where it was established that a relevant setup should consider the following aspects: i) the decoder only aims to return a list of messages without recovering users’ identities; ii) the error event should be defined per user; iii) the error probability has to be averaged over the users; iv) each user sends a fixed amount of information bits within a finite frame length. 

This formulation is well suited for short-packet random-access wireless communications since, in theory, it does not require coordination among users. As such, it captures the requirements of \gls{mmtc}, one of the new emerging communication scenarios in next generation wireless networks, where a huge amount of battery-limited devices is expected to connect sporadically to the network to send short information packets. Since its inception, this problem has been commonly referred in the literature as \gls{ura}. Several papers establishing fundamental limits for different relevant multiple-access channel models and setups appeared since then (see, e.g.,~\cite{Zadik19, Kowshik21, Ngo22, Ravi22}), and many transmission schemes trying to perform as close as possible to this fundamental limits has been proposed (e.g.,~\cite{Pra2021a,Ama2022,Tru2021a}).

The A-channel played an important role for codes design
in URA. In \cite{Ama2020a}, a coding scheme for AWGN \gls{ura} termed \gls{ccs} was introduced. It used a random inner code of size $q$ concatenated with an outer $q$-ary A-channel code. The A-channel code constructed for this purpose was termed \emph{tree code}. The flexibility of this code construction allowed it to be extended to different channel models. Several follow-up works on URA (e.g, \cite{Fen2021c,And2021,Lia2021,Ama2022,Che2022}) made use of an outer A-channel code. In \cite{Fengler21-05} an asymptotically Bayesian optimal inner decoder for the AWGN channel was constructed and it was shown that the \gls{ccs} construction can achieve the Shannon limit when $K$ and $q$ grow but its ratio $\lambda \to 0$. However, in practical applications the density $\lambda$ is not zero.

The A-channel is of relevance to URA in a more general sense: \emph{Every unsourced $K$-user code for $B$ bits at blocklength $n_0$ can be extended to a code of length $nn_0$ for $nBR_A(n, K)$ bits by concatenating with an outer unsourced A-channel code of rate $R_A$ with $n$ A-channel uses.} The loss in rate of $R_A$ does not appear in classical multiple-access where user identification is done based on the codebook. A system that can transmit 1 bit for each user with zero error can be used to transmit arbitrary many bits by simple repetition. For the unsourced channel this is not possible and an outer A-channel code is necessary to couple repeated transmissions.

Furthermore, the blocklength of the outer A-channel used for concatenated coding (e.g., \cite{Ama2020a,Fengler21-05,Fen2021c}) is in the order of $10-40$. Therefore, the asymptotic results for the A-channel are not necessarily insightful for code design.  

In this paper, we study the unsourced A-channel in the finite blocklength regime with arbitrary $K$ and $q$. In particular, we present two novel non-asymptotic achievability bounds. Also, we provide a second-order asymptotic approximation whose relevance is validated by means of numerical examples in different scenarios of interest. 

The A-channel finds interesting applications in noiseless non-adaptive group testing. The goal in group testing is to identify $K$ defective items in a large population of $N$ items by applying $T$ binary tests. A group-testing design is a $T\times N$ binary matrix where each column specifies the test in which that item participates. A test is declared positive if at least one tested item is defective.
Group testing was developed by Dorfman in 1943 \cite{Dor1943} for syphilis testing. Dorfman discovered that it is possible to test more people with a limited number of tests by pooling blood samples together. The topic has seen a recent rise in popularity since the COVID-19 pandemic led to a shortage of available tests for which group testing provides an appropriate solution. Group testing finds further important applications in DNA screening, large scale manufacturing control, neighborhood discovery, random access, machine learning, anomaly detection in routing networks, etc. \cite{Du2006,Ama2019a,Ber1984a,Mal2013,Xua2010}. For a recent survey on group testing from an information theoretic view, see \cite{Ald2019a}. 

The connection to the A-channel is as follows:
Each codebook for the unsourced $q$-ary A-channel with blocklength $n$ and size $M$ gives rise to a group-testing design for $N=M$ items with $T=nq$ tests. To convert the codebook to a group-testing design, each $q$-ary symbol $c_i$ is converted to a binary vector of size $q$ with a 1 at position $c_i$. The defective items take the role of the transmitting users and the set of defective items can be obtained by recovering the transmitted messages. 
This A-channel group-testing design has a fixed number $n$ of tests per item. The pair $(n,q)$ can be used to optimize the group-testing design. 

It is known that a fixed number of tests leads to improved error probabilities compared to an \gls{iid} Bernoulli test design, even if the average number of tests is the same \cite{Ald2019a}. A popular design, analyzed in \cite{Joh2019}, uses a fixed number of tests per item, which are chosen at random from all tests. Compared to that, an A-channel design offers more structure as each item participates in exactly one of each group of $q$ tests.
The Kautz-Singleton (KS) construction \cite{Kau1964} is another popular group-testing design that naturally has a $q$-ary structure. In particular, it is based on a $q$-ary Reed-Solomon code of length $n$. The KS construction was recently shown to be optimal for probabilistic group testing in certain scaling regimes \cite{Ina2018}. The random coding bound developed in this paper gives a concrete finite blocklength achievability result for a random, but highly structured, group-testing design. 

Motivated by the insights of our results and the algorithms developed in group testing, we also propose an improved decoder for the tree code.  Numerical simulations confirm that the improved decoder significantly increases the achievable rates of the tree code.




\section{Finite-Blocklength Framework}\label{sec:FBL}
%
%
We consider the channel model introduced in~\eqref{eq:channel1}, where $K$ users transmit codewords from a common codebook with entries drawn from a $q$-ary input alphabet $[q] =\{1,\ldots,q\}$ over $n$ channel uses of a noiseless channel. To denote the $n$-length input-output relation, we shall also write
\begin{equation}\label{eq:channel}
    \randvecy = \bigcup_{j\in[K]}\vecc_{j}
\end{equation}
where $\vecc_j\in [q]^n$ denotes the codeword transmitted by user $j$.  
We next define the notion of \gls{ura} code for the A-channel.
%
%
\begin{definition}[Code] \label{def:code}
    Let $\binom{[a]}{b}$ denote the set of combinations of $b$-element subsets of $[a]$. 
    Assume $q>K$, and let $W_j$, $j\in[K]$, denote the transmitted message by user $j$. An $(M,n,\epsilon)$-code for the unsourced A-channel~\eqref{eq:channel}, where $\vecc_j\in [q]^n$, consists of an encoder-decoder pair,
\begin{itemize}
    \item encoder: $f:[M]\mapsto [q]^n$;
    \item decoder: $g:\lrbo{\bigcup_{k=1}^{K}{[q]\choose k}}^n\mapsto {[M] \choose K}$, 
\end{itemize}
satisfying either the \gls{pupe} 
\begin{equation}\label{eq:pupe_def}
    P\sub{e}^{\rm (p)} \triangleq \prob{\{W_j \notin g(\randvecy)\} \cup  \{W_j = W_i, j\neq i\}} \le \epsilon
\end{equation}
or the \gls{jpe}
\begin{equation}\label{eq:jpe_def}
    P\sub{e}^{\rm (j)} \triangleq \prob{\{\{W_j\}_{j=1}^{K} \neq g(\randvecy)\} \cup  \{W_j = W_i, j\neq i\}} \le \epsilon.
\end{equation}
We assume that $\{W_j\}_{j=1}^{K}$ are independent and uniformly distributed on $[M]$, and that $f(W_j) = \vecc_j\in [q]^n$. For each type of error probability, we say the code achieves a \emph{rate} $R=\log_2 M/n$.
\end{definition}

Hence, we have $K$ users selecting randomly a codeword from a common codebook, and the decoder's task is to provide an estimate of the transmitted list of length $K$. In this paper, we assume $K$ is known at the receiver.
\subsection{Achievability Non-Asymptotic Bounds}\label{sec:ach}
In this section, we present our finite-blocklength achievability bounds for the unsourced A-channel. To do so, we consider a random-coding scheme where a codebook $\setC$ contains $M$ randomly generated codewords of length $n$ distributed according to $P_{\randvecx}(\vecc) = \prod_{i=1}^n P_{X}(c_i)$, where $P_X = {\rm Unif}[q]$. 
According to Definition~\ref{def:code}, user $j$ selects uniformly at random a message $W_j \in [M]$, and transmits the corresponding encoded codeword $f(W_j)= \vecc_{j}$. Due to symmetry, we assume without loss of generality that the first $K$ codewords are transmitted.  
We shall consider two different decoders, which will lead to our two different achievability bounds:
\paragraph*{\textbf{Cover decoder}} From the received sequence $\randvecy$, the decoder first discards all codewords from the codebook that are incompatible with the received sequence, i.e., those ones that are not covered by $\randvecy$. Then, the decoder outputs a list of $K$ codewords chosen uniformly at random from the surviving codewords. 
Since the A-channel is noiseless, the list of surviving codewords always contains the transmitted list plus $N_{\rm fa,c} \in[0:M-K]$ false alarms. Therefore, $P\sub{e}^{\rm (p)}$ can be upper-bounded by the \gls{pupe} achieved by this decoding rule, namely, $P\sub{e}^{\rm (p)}\leq \Ex{N\sub{fa,c}}{\frac{N\sub{fa,c}}{K+N\sub{fa,c}}}$. 
Similarly, $P\sub{e}^{\rm (j)}$ can be upper-bounded by the probability of having at least one false alarm, i.e., $P\sub{e}^{\rm (j)}\leq \prob{N\sub{fa,c}\geq 1}$.  
\paragraph*{\textbf{Joint decoder}} This decoder finds all combinations of $K$ codewords from the codebook that can be selected to generate the output $\randvecy$. 
If there is more than one valid combination, the decoder chooses one, uniformly at random, and outputs the list of indices in that combination. Note, that this is exactly the maximum likelihood decoder.
Since the A-channel is noiseless, the combination containing only the $K$ transmitted codewords will always be valid.  A wrong combination will differ from the correct one in $N\sub{fa,j} = N\sub{md,j}$ indices, i.e., same number of misdetections and false alarms. Hence, we can bound the error probability as $P\sub{e}^{\rm (p)}\leq \Ex{N\sub{fa,j}}{\frac{N\sub{fa,j}}{K}}$ and $P\sub{e}^{\rm (j)}\leq \prob{N\sub{fa,j}\geq 1}$.   
\begin{remark}
    Recall that, in this paper, we assumed $K$ to be known at the receiver. The cover decoder does not require this knowledge and works unaltered if $K$ is unknown. The joint decoder can be adopted in two ways to deal with the missing information. One possibility is to extend the code design and use additional channel uses to estimate the number of users. Another way is to let the receiver find the smallest set of messages that recreate the channel output, as in the smallest satisfying set algorithm in group testing \cite{Ald2019a}.
\end{remark}
We are now ready to present our two achievability bounds.
\begin{theorem}[Cover decoding]\label{th:ach_A_ch}
There exists an $(M,n,\epsilon)$-code for the unsourced $K$-user A-channel with \gls{pupe} satisfying
\begin{IEEEeqnarray}{lCl}
    \epsilon &&\leq \sum_{\ell=1}^{K-1}\frac{\ell}{K+\ell}\mathbb{E}\Biggl[\min\Biggl\{1,\binom{M-K}{\ell} \prod_{k=1}^{K}\lro{\frac{k}{q}}^{A_k\ell}\Biggr\}\Biggr] \nonumber\\
    &&{} + \mathbb{E}\Biggl[\min\Biggl\{1,\binom{M-K}{K} \prod_{k=1}^{K}\lro{\frac{k}{q}}^{A_k K}\Biggr\}\Biggr] + \frac{\binom{K}{2}}{M}\label{eq:error_UB}
\end{IEEEeqnarray}
and there exists an $(M,n,\epsilon)$-code with \gls{jpe} satisfying
\begin{IEEEeqnarray}{lCl}
    \epsilon &&\leq \frac{\binom{K}{2}}{M} + \mathbb{E}\Biggl[\min\Biggl\{1,(M-K) \prod_{k=1}^{K}\lro{\frac{k}{q}}^{A_k}\Biggr\}\Biggr]. \IEEEeqnarraynumspace \label{eq:error_UB_jpe}
\end{IEEEeqnarray}
In both \eqref{eq:error_UB} and \eqref{eq:error_UB_jpe}, $A_k$ is the $k$-th element of $\randveca = \tp{[A_1, \dots, A_K]}$, which is a multinomial-distributed random vector with $n$ trials and $K$ possible outcomes with probabilities $\{p_k\}_{k=1}^K$, which are given by 
\begin{equation}
    p_k = \frac{q!S(K,k)}{(q-k)! q^K}  \label{eq:stirling}
\end{equation}
where $S(K,k)$ denotes the Stirling number of the second kind \cite[Sec. 26.8.6]{olver10}. 
\end{theorem}
\begin{IEEEproof}
See Appendix~\ref{app:proof_ach_A_ch}.
\end{IEEEproof}
\begin{theorem}[Joint decoding]\label{th:ach_A_ch_joint}
There exists an $(M,n,\epsilon)$-code for the unsourced $K$-user A-channel with \gls{pupe} satisfying
\begin{IEEEeqnarray}{lCl}
    \epsilon
    &\leq& \frac{\binom{K}{2}}{M} + \sum_{\ell=1}^{K} \frac{\ell}{K}\mathbb{E}\Biggl[\min\Biggl\{1,\binom{K}{K-\ell} \binom{M-K}{\ell }\nonumber\\
    &&\qquad\qquad\qquad\qquad \times \prod\limits_{k=1}^{K}\Biggl(\sum\limits_{\eta=\underline{\eta}}^{\overline{\eta}}\bar{p}_\eta p(k,\ell,\eta)\Biggr)^{A_k}\Biggr\}\Biggr] \IEEEeqnarraynumspace \label{eq:error_UB_pupe}
\end{IEEEeqnarray}
and there exists an $(M,n,\epsilon)$-code with \gls{jpe} satisfying
\begin{IEEEeqnarray}{lCl}
    \epsilon &\leq& \frac{\binom{K}{2}}{M} + \sum_{\ell=1}^{K}\mathbb{E}\Biggl[\min\Biggl\{1,\binom{K}{K-\ell} \binom{M-K}{\ell }\nonumber\\
    &&\qquad\qquad\qquad\qquad \times \prod\limits_{k=1}^{K}\Biggl(\sum\limits_{\eta=\underline{\eta}}^{\overline{\eta}}\bar{p}_\eta p(k,\ell,\eta)\Biggr)^{A_k}\Biggr\}\Biggr]. \IEEEeqnarraynumspace \label{eq:error_UB_joint}
\end{IEEEeqnarray}
In~\eqref{eq:error_UB_pupe} and~\eqref{eq:error_UB_joint},
\begin{equation}
    \bar{p}_\eta = \frac{k! S(K-\ell,\eta)}{(k-\eta)!  k^{K-\ell} Z_\eta}
\end{equation}
with $Z_\eta$ being a normalizing constant ensuring that $\sum_{\eta=\underline{\eta}}^{\overline{\eta}} \bar{p}_\eta = 1$. Here $\underline{\eta} \triangleq \max\{0,k-\ell\}$ and $\overline{\eta} \triangleq \min\{k,K-\ell\}$. Finally
\begin{equation}
p(k,\ell,\eta) = \lro{\frac{k}{q}}^\ell\pi(k,\ell,\eta)  \label{eq:p_coupon}  
\end{equation}
where the first factor $(k/q)^\ell$ is the probability that the $\ell$
non-transmitted codewords hit one of the $k$ output symbols, and $\pi(k,\ell,\eta)$ is the conditional probability that the $\ell$ non-transmitted codewords
hit the remaining $k-\eta$ symbols given they all hit one of the $k$ output symbols. Note that the probability $\pi(k,\ell,\eta)$ resembles the classical coupon collector problem, which can be modelled by the Markov chain depicted in Fig.~\ref{fig:markov_chain}. Specifically, the problem is analogous to the coupon collector problem in the sense that $\pi(k,\ell,\eta)$ is the probability of collecting $k$ out of $k$ possible coupons in $\ell$
steps when starting with $\eta$ coupons. The case $\eta=0$ can be computed in closed form as $\pi(k,\ell,0) = S(\ell,k)k!/k^\ell$.
\begin{figure}[t!]
        \centering
        \includegraphics[width=\columnwidth]{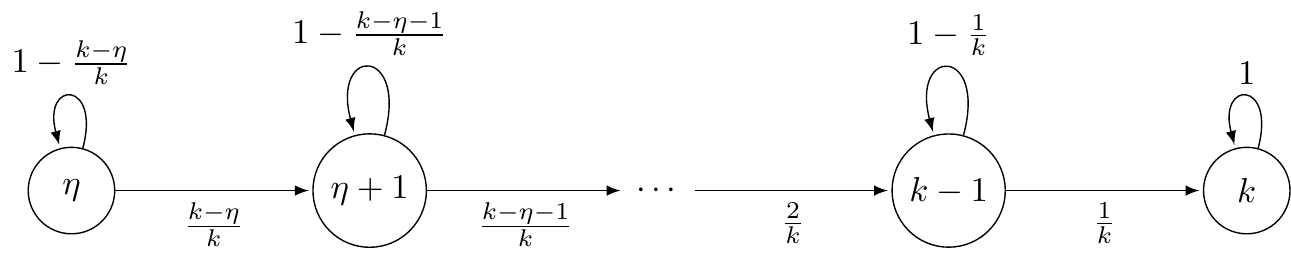}
        \caption{Markov chain describing the state evolution yielding $\pi(k,\ell,\eta)$, which denotes the probability that $\ell$ non-transmitted codewords hit the remaining $k-\eta$ symbols of $Y_i$ of cardinality $k$ at channel use $i$, conditioned on the fact that the the $\ell$ non-transmitted codewords lie within the set of symbols in $Y_i$.}
        \label{fig:markov_chain}
    \end{figure}
For $\eta>0$, $\pi(k,\ell,\eta)$ can be efficiently computed recursively. The specific formulas are given in Appendix~\ref{app:coupon_gen_eta}.
\end{theorem}
\begin{IEEEproof}
See Appendix~\ref{app:proof_ach_A_ch_joint}.
\end{IEEEproof}
\subsection{Asymptotic Analysis}\label{sec:asymp}
Let 
\begin{IEEEeqnarray}{lCl}
\mu_\ell(K,q) &\triangleq& \Ex{}{\log \frac{P_{Y|\randvecx_{[K]}}(Y|\randvecx_{[K]})}{P_{Y|\randvecx_{[K-\ell]}}(Y|\randvecx_{[K-\ell]})}}\nonumber\\
&=& I(\randvecx_{[K-\ell+1:K]};Y|\randvecx_{[K-\ell]}),\label{eq:MI}\\
\sigma^2_\ell(K,q) &\triangleq& \Var{\log \frac{P_{Y|\randvecx_{[K]}}(Y|\randvecx_{[K]})}{P_{Y|\randvecx_{[K-\ell]}}(Y|\randvecx_{[K-\ell]})}}\label{eq:DISP}
\end{IEEEeqnarray}
where $\randvecx_{S} = (X_i)_{i\in S}$ for any $S\subset [K]$. 
We drop the explicit dependence on $K,q$ for readability whenever it is clear from the context, so $\mu_\ell \equiv \mu_\ell(K,q)$. Since the channel is noiseless, 
$\mu_K = -\mathbb{E}[\log P_Y(Y)]$, i.e., the mutual information coincides with the output entropy, and $\sigma^2_\ell = -\Var{\log P_Y(Y)}$.\footnote{When $P_Y$ is the output distribution induced by a capacity achieving input distribution, $\mu_K$ is also the channel capacity.}
In the case $\ell=K$, each output sequence $y$ with cardinality $k$ 
has probability
$P_{Y,|Y|}(y,k) = S(K,k)k!/q^K$.
Since there are $\binom{q}{k}$ different outputs $y$ for $|y| = k$,
\begin{IEEEeqnarray}{lCl}
    \mu_K(K,q) 
    &=& -\sum_{k=1}^K \sum_{y:|y|=k} \frac{S(K,k)k!}{q^K}\log\frac{S(K,k)k!}{q^K}\IEEEeqnarraynumspace\\
    &=& -\sum_{k=1}^K \binom{q}{k}\frac{S(K,k)k!}{q^K} \log\frac{S(K,k)k!}{q^K}\\
    &=& K\log q - \sum_{k=1}^K p_k \log (S(K,k)k!) \label{eq:entr_out}
\end{IEEEeqnarray}
where in the last equality we used the definition of $p_k$ 
 in~\eqref{eq:stirling}. The output entropy for the noiseless A-channel with uniform inputs \eqref{eq:entr_out} was already obtained in \cite{Chang81,Bassalygo2000}.
By similar steps,
\begin{IEEEeqnarray}{lCl}
    \sigma^2_K(K,q) 
    &=& \sum_{k=1}^K p_k\left[\log\frac{S(K,k)k!}{q^K}\right]^2-\mu_K(K,q)^2.\IEEEeqnarraynumspace
    \label{eq:var}
\end{IEEEeqnarray}
Throughout the rest of this section, we will use
\begin{IEEEeqnarray}{lCl}
    I_{K,q} &\triangleq& \frac{\mu_K(K,q)}{K\log q},\label{eq:I_last}\\
    V_{K,q} &\triangleq& \frac{\sigma^2_K(K,q)}{(K\log q)^2}.\IEEEeqnarraynumspace
    \label{eq:V_last}
\end{IEEEeqnarray}

Recall that in Theorem~\ref{th:ach_A_ch_joint}, $A_k$ is the $k$-th entry of $\randveca = \tp{[A_1,\dots,A_k]}$, which is multinomial distributed with parameters $\{p_k\}_{k=1}^K$ (with $p_k$ given in \eqref{eq:stirling}) and $\sum_{k=1}^K A_k = n$. Let $c_k \triangleq \log_2 (\sum_{\underline{\eta}}^{\overline{\eta}} \bar{p}_\eta p(k,\ell,\eta))$. It follows that $\sum_{k=1}^K A_k c_k \eqdist \sum_{i=1}^n Z_i$, where $\eqdist$ denotes equality in distribution, and where $\{Z_i\}_{i=1}^n$ is a sequence of \gls{iid} random variables taking values on $c_k$ with probability $p_k$ for $k\in[K]$. In the following, a generic realization of the random variable $Z_i$ will be denoted simply by $Z$. Then, by applying the so-called normal approximation (Berry-Esseen theorem~\cite[Ch.~XVI.5]{feller71-a} and~\cite[Lemma~47]{polyanskiy10-05a}) to the expected value of~\eqref{eq:error_UB_pupe}, it follows that for some constant $B$ independent of $n$ (see, e.g., \cite[Eqs.~(255)-(267)]{polyanskiy10-05a}),
\begin{IEEEeqnarray}{lCl}
    \epsilon &\leq& \sum_{\ell=1}^{K} \frac{\ell}{K}Q\lro{\frac{\frac{\mu_{\ell}}{\ell} - R \log_2 q - \frac{\log_2\lro{ (\frac{e}{\ell})^\ell \binom{K}{K-\ell}} }{\ell n}}{ \sigma_{\ell}/(\ell\sqrt{n})}}  \nonumber \\
    &&{}+\frac{B}{\sqrt{n}} + \frac{\binom{K}{2}}{M}. \label{eq:NA_Q_funcs}
\end{IEEEeqnarray}

It is shown in Appendix \ref{sec:MI_ineq_proof} that basic properties of the conditional mutual information and the symmetry of the $X_i$'s imply 
\begin{equation}
    \frac{\mu_\ell}{\ell} \geq \frac{\mu_{\ell+1}}{\ell+1}
    \label{eq:MI_ineq}
\end{equation} 
for every $\ell\in[K-1]$. Then, as $n$ grows and the rate approaches $I_{K,q}$, the $\ell = K$ term in \eqref{eq:error_UB_pupe} becomes dominant while the $\ell<K$ terms still decay exponentially fast with $n$.
\begin{remark}
Usually, the capacity region of the multiple access channel is the union of $K$-dimensional pentagon constrained by the different conditional mutual information terms $\mu_\ell$. In the unsourced case, where all input distributions are constrained to be equal, it is apparent from \eqref{eq:MI_ineq} that $\mu_K$ is the most constraining limit and therefore it dominates the $n\to \infty$ limit. Equation \eqref{eq:NA_Q_funcs} shows that the conditional mutual information terms still influence the random coding error probabilities in the finite blocklength regime. Nonetheless, their contribution vanishes exponentially with the blocklength.
\end{remark}

\begin{figure}[t!]
    \centering
    \begin{subfigure}{\columnwidth}
    \centering
    \includegraphics[width=\columnwidth]{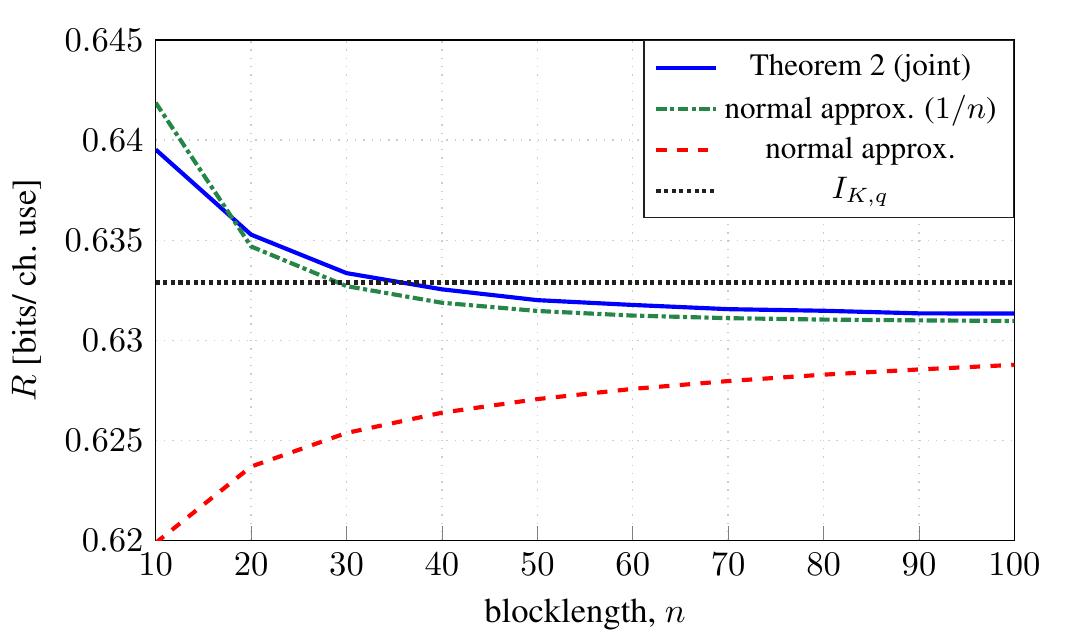}
    \caption{$10\leq n \leq 100$.}
    \label{fig:NA_small_n}
\end{subfigure}

\begin{subfigure}{\columnwidth}
    \centering
    \includegraphics[width=\columnwidth]{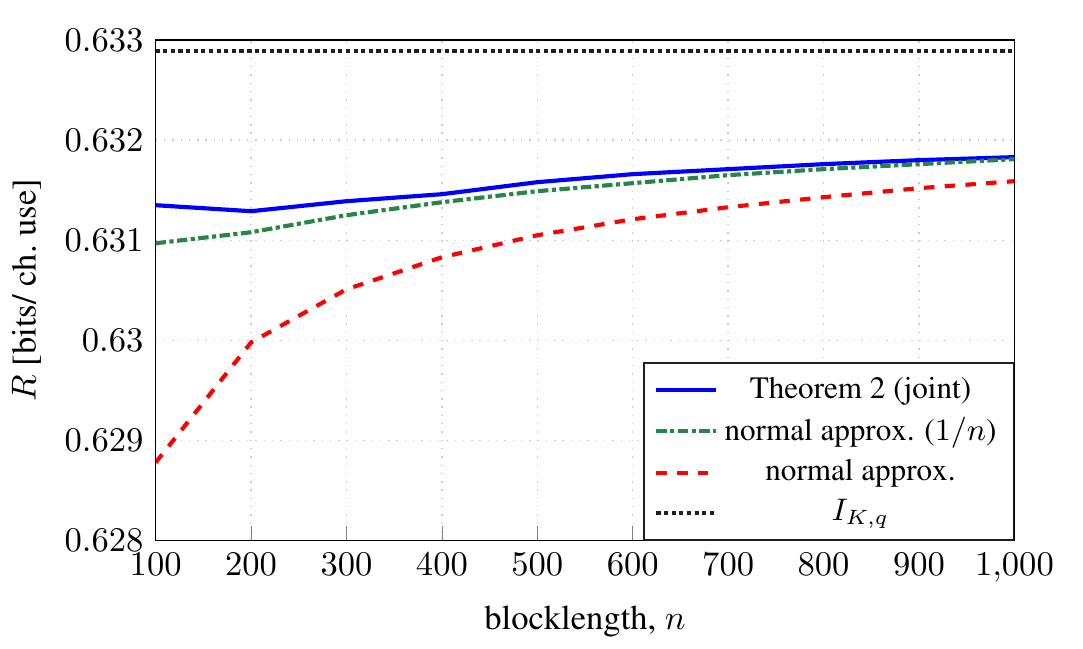}
    \caption{$100 \leq n \leq 1000$.}
    \label{fig:NA_large_n}
\end{subfigure}
\caption{Rate versus blocklength $n$ for $q=16$, $K=5$ and $\epsilon = 0.05$.}
\label{fig:NA}
\end{figure}

By collecting the $\ell < K$ terms and $\binom{K}{2}/M$ in \eqref{eq:NA_Q_funcs} in a $o(1/n)$ term, after some standard manipulations, \eqref{eq:NA_Q_funcs} can be expressed in terms of the rate as
\begin{equation}
    R = I_{K,q} - \sqrt{\frac{V_{K,q}}{n}}Q^{-1}\lro{\epsilon - \frac{B}{\sqrt{n}} + o\lro{\frac{1}{n}}} + \frac{\log_2(K/e)}{n\log_2(q)}.
    \label{eq:na_naive}
\end{equation}
The constant $B$ is determined by the Berry-Esseen theorem~\cite[Ch.~XVI.5]{feller71-a} and~\cite[Lemma~47]{polyanskiy10-05a}. For sufficiently large $n$, it follows that $Q^{-1}(\epsilon -B/\sqrt{n} + o(n^{-1})) = Q^{-1}(\epsilon) + \overline{B}/\sqrt{n} + \mathcal{O}(1/n)$, for some $\overline{B}$ independent of $n$. Numerical experiments suggest that for the A-channel, the value of $B$ that can be obtained by applying the Berry-Esseen theorem~\cite[Ch.~XVI.5]{feller71-a} and~\cite[Lemma~47]{polyanskiy10-05a} is not tight. In other words, $\sum_{i=1}^n Z_i$ converges much faster to the Gaussian distribution than the speed suggested by $B$. 
In Fig.~\ref{fig:NA}, we show that the approximation 
\begin{equation}
    R = I_{K,q} - \sqrt{\frac{V_{K,q}}{n}}Q^{-1}\lro{\epsilon} + \frac{\log_2(K/e)}{n\log_2(q)}
    \label{eq:na_improved}
\end{equation}
can indeed provide accurate estimates of the bound provided in Theorem~\ref{th:ach_A_ch_joint} for small values of $n$. This approximation is tight as long as the true value of $B$ is sufficiently small so that the resulting $\overline{B}$ is much smaller than $\log_2(K/e)/\log_2(q)$. When this is true, ignoring the term $\overline{B}/{\sqrt{n}}$ does not compromise the accuracy of the approximation for small $n$.

This is shown in Fig.~\ref{fig:NA}, where we compare the non-asymptotic random coding bound with joint decoding given in Theorem~\ref{th:ach_A_ch_joint}, and the normal approximation \eqref{eq:na_improved} with and without the $1/n$-term. We further plot the maximum coding rate achievabile with uniform inputs $I_{K,q}$. We can observe that the $1/n$-term of the normal approximation is necessary to capture the behaviour of the non-asymptotic bound in the small blocklength regime, where rates are higher than $I_{K,q}$ (Fig.~\ref{fig:NA_small_n}). As $n$ grows large the dispersion term becomes dominant and the achievability curve starts to show the typical $1/\sqrt{n}$ convergence to the asymptotic limit from below (Fig.~\ref{fig:NA_large_n}).

\section{A-Channel Code: Tree Code}\label{sec:code}
%
\subsection{Code Construction}\label{sec:tree_code}
%

A $B$-bit message is divided into blocks of size 
$\{b_i\}_{i=1}^n$ such that $\sum_{i=1}^n b_i = B$ and such that
$b_1 =  J$ and $b_i < J$ for all $i \in [2:n]$. 
Each subblock $i \in [2:n]$ is augmented to size $J$ by appending
$\pi_i = J - b_i$ parity bits,  
obtained using pseudo-random linear combinations of the information bits of the
previous blocks $i' < i$.
Note that there is a one-to-one association between the set of all sequences of coded blocks and
the paths of a tree of depth $n$.
The pseudo-random parity-check equations generating the parity bits 
are identical for all users, i.e., each user makes use exactly of the same outer {\em tree code}. This makes the code compatible with
the unsourced paradigm. Each user then transmits the $n$ coded symbols
over the $2^J$-ary A-channel.

Let $Y_i,\ i\in[n]$,
be the channel outputs of the A-channel. 
Since the sections contain parity bits with parity profile $\{0,\pi_2, \ldots, \pi_n\}$, 
not all message sequences in $Y_1 \times Y_2 \times \cdots \times Y_n$ are possible. 
The role of the outer decoder is to identify all possible message sequences, i.e., those corresponding to
paths in the tree of the outer tree code \cite{Ama2020a}.  
The output list $\Lc$ is initialized as an empty list. Starting from $i=1$ and proceeding in order, the decoder converts all the integer indices in 
$Y_i$ back to their binary representation, separates data and parity bits, computes the parity 
checks for all the combinations with messages from the list $\Lc$, and extends only the paths 
in the tree which fulfill the parity checks.
A precise analysis of the error probability
in various  asymptotic regimes as well as an algorithm to optimize
the parity profile for a target complexity and error probability are provided in \cite{Ama2020a}.

The analysis in \cite{Ama2020a} and \cite{Fengler21-05} showed that the tree code performs well 
in the regime of vanishing sparsity, i.e., $K/q \to 0$, which is the regime where both joint and cover decoding bounds (see Theorems~\ref{th:ach_A_ch} and~\ref{th:ach_A_ch_joint}) perform similarly. However, for moderate sparsity, our numerical evaluation of Theorems~\ref{th:ach_A_ch} and~\ref{th:ach_A_ch_joint} reveals that the joint and cover decoding bounds exhibit a considerable gap (See Fig.~\ref{fig:tree_code}).  Since the original tree decoder outputs all codewords that satisfy the parity checks, the tree code described above cannot outperform the cover decoding bound. In the next section, we propose enhanced decoding strategies for the original tree code based on ideas from group testing and insights from the analysis of the joint decoder.  
 
 \subsection{Enhanced Decoding}\label{sec:enh_decoding}
 The proof of Theorem \ref{th:ach_A_ch_joint} shows that joint decoding can improve upon cover decoding by considering combinations of codewords instead of just individual codewords.
 In this section, we use this concept to develop two improved decoding algorithms for the tree code. These methods
 strictly improve the performance of the tree code since they consists of a post-processing step of the output list when the output list is greater than $K$. In earlier works such as \cite{Fen2021c} and \cite{Fengler21-05},
 codewords were discarded at random to reduce the output list to the required size. This necessarily results in a large number of errors when the output list is significantly larger than $K$. Since the output list contains only false alarms and no misdetections, the decoding performance can be improved by filtering the output list to remove false alarms. 
 Let the size of the cover decoder output list be 
 $K + \Delta$. A valid strategy is to check all
 ${K + \Delta \choose K}$ combinations of $K$ codewords from the list. Of course this leads to a complexity that grows exponentially in $\Delta$. When $K$ is not known, one can search for the combination with the least codewords that produces the channel output. In the group-testing literature, this approach is called the smallest-satisfying set (SSS) method \cite{Ald2019a}. Note that finding the SSS is in general NP hard, as it can be shown to be
 equivalent to the set cover problem \cite[Remark 2.1]{Ald2019a}. In the following we describe two methods, developed for group testing, that approximate the combinatorial search in a greedy manner. In particular, we will consider the so-called
 definitive defectives (DD) and sequential combinatorial pursuit (SCOMP) algorithms~\cite{Ald2014}.\footnote{An alternative approach is based on linear programming \cite{Mal2012}. It is very similar to SCOMP in terms of achievable rates and complexity, so we exclude it from the comparison in this paper. A more detailed comparison is left for future work.}
They both work by filtering the original output list. Specifically, SCOMP is a strict improvement over DD, in the sense that it consists of applying DD followed by an additional processing step. Therefore, the algorithm can be chosen based on complexity and/or rate requirements, since each processing step increases the decoding complexity, but also increases the performance. 
\paragraph*{\bf DD } As a first step we re-encode all the messages in the output list of the tree decoder, which we denote by $\mv_1,...,\mv_{|\Lc|}$. For $i \in [n]$, the DD algorithm isolates all indices $i$ for which $m_{j,i}$ is unique among $\{m_{1,i},...,m_{|\Lc|,i}\}$. The messages with indices isolated this way have for sure been transmitted since they were the only ones in the list that can explain the observed channel output. Let $\Lc_{\text{DD}}$ be the list of isolated messages and let $\Lc_R$ denote the remaining messages that were not isolated. If $|\Lc_\text{DD}|<K$ we choose random messages from $\Lc_R$ to fill the output list up to size $K$.  
\paragraph*{\bf SCOMP }
The SCOMP algorithm proceeds by scanning the list of remaining entries $\Lc_R$ after DD processing for
 			appropriate candidates using the following greedy heuristic:
 			i) The symbols in the channel output that have been covered by the DD list are removed. The remaining symbols are called \emph{unexplained}. ii) The index $j_\text{max}$ is searched for which $\mv_j$ covers the \emph{most} unexplained symbols. This index is added to  the output list $\Lc_\text{SCOMP} = \Lc_\text{DD}\cup j_\text{max}$. iii) The symbols covered by $\mv_{j_\text{max}}$ are removed from the list of unexplained symbols. The algorithm repeats this process until no unexplained symbols are left. If $|\Lc_\text{SCOMP}|<K$ we again add messages at random. This algorithm will always terminate in at most $K$ steps, since the transmitted messages are always contained in the original output list.

 \subsection{Numerical Results}  
 \begin{figure}[t!]
    \centering
    \begin{subfigure}{\columnwidth}
        \centering
        \includegraphics[width=\columnwidth]{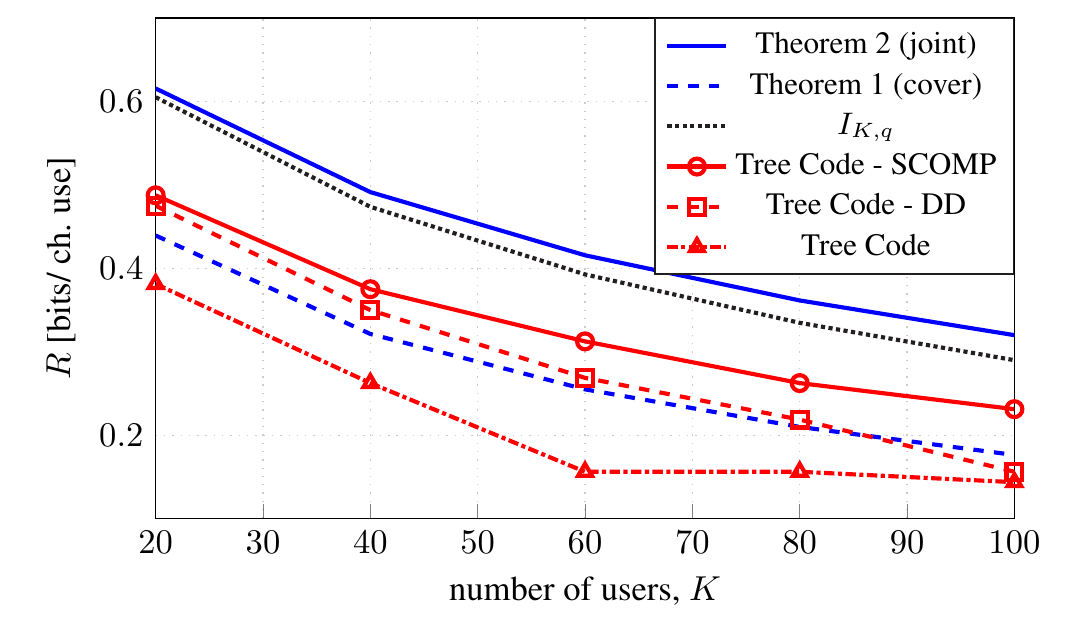}
        \caption{$J=8$, $n=20$, $\epsilon = 0.05$}
        \label{fig:tree_code_K}
\end{subfigure}

 \begin{subfigure}{\columnwidth}
        \centering
        \includegraphics[width=\columnwidth]{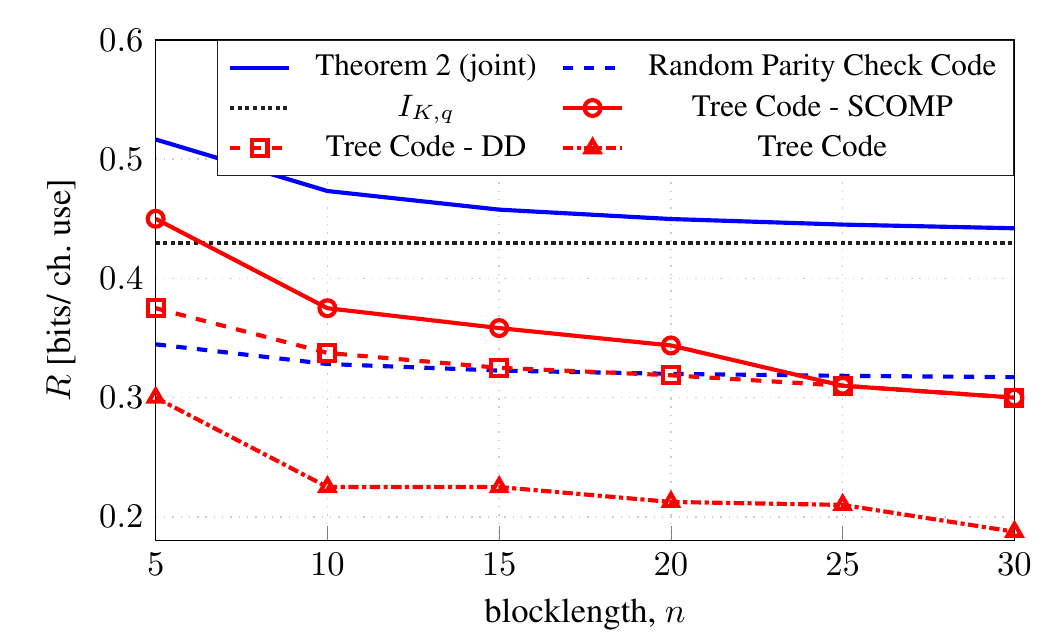}
        \caption{$J=8$, $K=50$, $\epsilon = 0.05$}
        \label{fig:tree_code_n}
\end{subfigure}
\caption{Rate versus number of active users $K$ (Fig.~\ref{fig:tree_code_K}) and versus blocklength $n$ (Fig.~\ref{fig:tree_code_n}) including the tree code with DD and SCOMP post-processing.}
\label{fig:tree_code}
\end{figure}
In Fig.~\ref{fig:tree_code}, we compare the performance of the original tree code described in Section~\ref{sec:tree_code}  with the enhanced versions described in Section~\ref{sec:enh_decoding}. As performance benchmarks, we use the finite-blocklength bounds derived in Theorems~\ref{th:ach_A_ch} (cover decoding) and~\ref{th:ach_A_ch_joint} (joint decoding), and the maximum coding rate achievabile asymptotically by uniform inputs $I_{K,q}$~\eqref{eq:I_last}. We use $q = 2^J$ with $J=8$. Let the rate $R = B/(B+P)$, where $B$ denotes the number of information bits, and $P$ the number of parity check bits. We fix the error constraint $\epsilon \leq 0.05$, and select the largest rate $R$ (smallest value of $P$) such that the error constraint is satisfied.
The parity profile is set by choosing $\pi_n = J$ and dividing the remaining parity check bits evenly between sections $2,...,n$. If the remaining parity check bits cannot be divided evenly, the later sections are prioritized. We remark that the resulting parity profile provides a good balance between decoding complexity and error probability. 

We can observe that there is a considerable gap between joint and cover decoding. Furthermore, we can observe that $I_{K,q}$ is exceeded for small blocklengths as are the achievable rates of all tree code variants. We can also observe that the suggested group-testing-motivated post-processing strategies (Tree code - DD, Tree code - SCOMP) allow to increase the achievable rates of the tree code significantly. Remarkably, both DD and SCOMP post-processing strategies allow to outperform the cover decoding bound. 

\section{A-Channel Designs in Group Testing}
Recall from Section~\ref{sec:intro} that an unsourced $q$-ary A-channel code of blocklength $n$
and size $M$ can be thought of as a group-testing matrix for $N=M$ items with $T=nq$ tests. Here, the number of active users is analog to the number of defective items. The tests are divided into $n$ groups of size $q$ so that each item participates in exactly $n$ tests, i.e., in one test per group.
Even though A-channel-based group-testing constructions are less flexible (they require the number of tests $T$ to be a multiple of $q$), they also provide more structure, which allows for efficient recovery and an easier analysis. 

The finite-blocklength achievability bounds given in Theorems~\ref{th:ach_A_ch} and~\ref{th:ach_A_ch_joint} allow to compute concrete achievable test numbers for a fixed $q$ 
and a fixed error probability $\epsilon$. In particular, $q$ can be seen as an optimization parameter that can be chosen to minimize the number of required tests. The analogy between unsourced A-channel codes and group testing motivates the following results.
\begin{corollary}
    There exist group-testing matrices, constructed from unsourced A-channel codes, such that $d$ defective items out of $N$ items can be recovered with $T = nq$ tests and the error probability given by Theorems~\ref{th:ach_A_ch} and~\ref{th:ach_A_ch_joint} without the penalty term $\binom{K}{2}/M$ (since random collisions among items are not possible). 
\end{corollary}

The following theorem shows that it is possible to achieve the optimal number of tests $T = \mathcal{O}(d\log N)$. 
\begin{theorem}
\label{th:gt_scaling}
There exists a sequence of group-testing matrices, constructed from unsourced A-channel codes, such that $d$ defective items out of $N$ items can be recovered with an error probability that vanishes in the limit $d,N,T \to \infty$ if
\beq
	T =  d\log N.
	\label{eq:T_scaling}
\eeq
\end{theorem} 
\begin{IEEEproof}
Let $K,q\to\infty$ with $\lambda = K/q$ fixed. The mutual information for the $K$-user A-channel with uniform inputs in this limit is given by~\cite{Bassalygo2000}:
\beq
	\lim_{K,q \to \infty} \frac{I(\randvecx_{[K]};Y)}{q} = h(1-e^{-\lambda})
\eeq 
where $h(\cdot)$ is the binary entropy function.
By the channel coding theorem~\cite[Ch. 7.7]{cover06-a}, there exist codes with sumrates $R_\text{sum} = K\log M/(nq)$ for which the error probability vanishes as long as $R_\text{sum} < h(1-e^{-\lambda})$. The right hand side is maximised for $\lambda = \ln 2$. Assuming that a code achieving this performance is used, we obtain \eqref{eq:T_scaling} by replacing $n = K\log M/q$ in $T = nq$, and using that in the standard group-testing notation $M=N$.
\end{IEEEproof}
If the optimal sparsity $\lambda = \ln 2$ cannot be attained, the number of required tests becomes $T = h(1- e^{-\lambda})^{-1}d\log N$.  
This result lies in the realm of probabilistic group testing \cite{Ald2019a} as for finite values of $N$ and $T$, there is
always a non-zero chance of failure, albeit it can be made arbitrary small by increasing $N$ and $T$.   
Note also that the relative scaling of $N$ and $d$ is not specified in Theorem~\ref{th:gt_scaling}. It is implicitly assumed though, through the order of limits (first $N,T\to\infty$ then $d,q \to \infty$), that $N$ is much larger than $d$. Instead of taking the second limit, we can generalize \eqref{eq:T_scaling} to hold for all $d$ by setting
\begin{equation}
T = d \log N \min_q \frac{q}{\mu_d(d,q)}
\end{equation}
where $\mu_d(d,q)$ is given in \eqref{eq:entr_out}. 

It is known that group-testing designs with a constant number $w$ of tests per item
perform better than unstructured random designs, even when the average number of test per item is the same \cite{Joh2019}. A commonly analyzed setup consists in fixing $w$ and choosing the tests randomly from the set of all $w$-weight vectors. An A-channel design also has a fixed number of test per item
$w = T/q$ but has even more structure, which provides some advantages. In particular, the $q$-ary structure of the A-channel allows to represent the group-testing matrix in an efficient way using only $n\log q$ bits to specify the test in which each item participates. If a structured code is used, such as the tree code, the group-testing matrix does not need to be stored explicitly as each column can be constructed in $\mathcal{O}(\log N)$ time.
Furthermore, the reconstruction of the defective set can be done in $\mathcal{O}(d^2\log N)$ time. As such, the tree code falls into the category of sub-linear group-testing
designs \cite{Ald2019a}. They are especially useful in problems where the recovery time
is the limiting factor, rather than the acquisition of tests. This is the case, for example, in big data and computer science applications. Theorem \ref{th:gt_scaling} shows the existence of A-channel codes achieving the optimal test scaling, but it requires $q$ to scale proportional to $d$. The analysis of the tree code in such a scaling regime is an interesting open problem, which is left for future work.
\subsection{Numerical Results}\label{sec:numeric}
\begin{figure}[t!]
    \centering
    \includegraphics[width=\columnwidth]{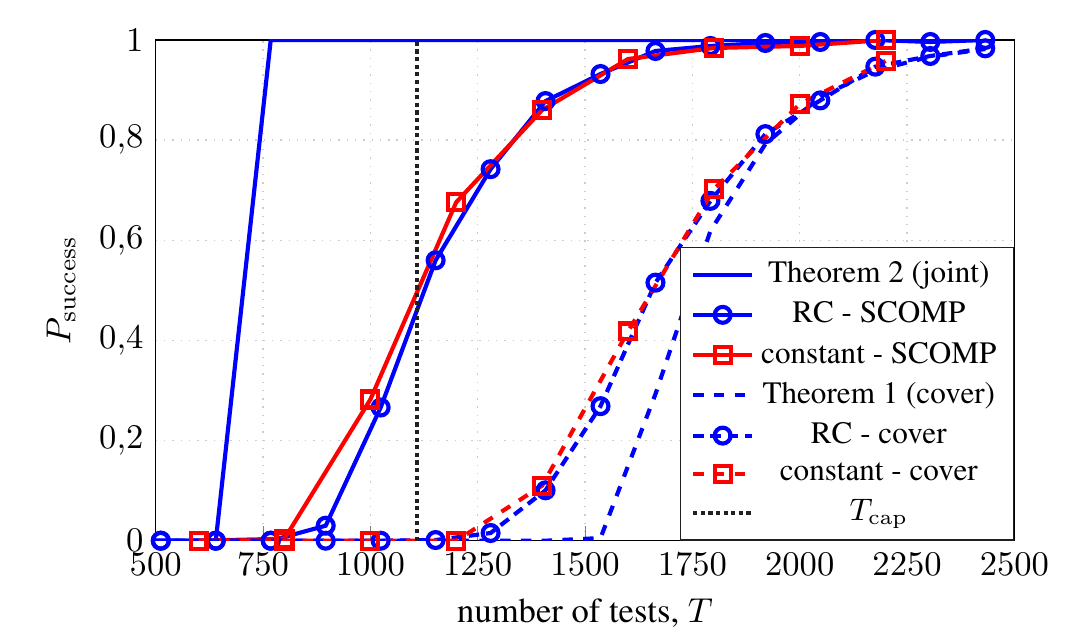}
    \caption{$P_{\rm success}$ versus number of tests $T$ in a group-testing setup with $d=100$ and $N=2000$. In the A-channel designs, $q=2^7$. }
    \label{fig:group_testing}
\end{figure}
Fig.~\ref{fig:group_testing} shows the performance of Theorems~\ref{th:ach_A_ch} and~\ref{th:ach_A_ch_joint} (without the $\binom{K}{2}/M$ term) in the group-testing setup in terms of probability of success $P_\text{success} = 1 - P_e^{({\rm j})}$, where a success is declared if the set
of defective items is perfectly recovered.\footnote{This corresponds to probabilistic group testing. For the PUPE bounds in Theorems~\ref{th:ach_A_ch} and~\ref{th:ach_A_ch_joint}, this would correspond to partial recovery \cite{Ald2019a}[Ch. 5.1].}    
We compare our achievability bounds
with empirical error rates achieved by a random A-channel code under
cover (RC - cover) and SCOMP (RC - SCOMP) decoding. The black dotted line shows $T_\text{cap} \triangleq d \log N q/\mu_d(d,q)$ for $q=2^7$, which provides an asymptotic achievability bound since for $T>T_\text{cap}$, by the channel coding theorem, there exist A-channel codes achieving $P_\text{success} \to 1$ in the limit $N,T\to\infty$ with $(\log N)/T$ fixed.  We also compare our bounds
with a constant design (constant - cover; constant - SCOMP in Fig.~\ref{fig:group_testing}) with exactly $w = (\ln 2) T/d$ test per item given in \cite{Ald2019a}. 

We assume that $d$ out of $N$ items are defective and set $d=100$ and $N=2000$. 
We choose $q=2^7$, which was found empirically to give the best results.  As we can observe, the A-channel design, which has $w = T/q$, exhibits almost the same performance as the constant weight designs, when using both the cover and the SCOMP decoders.  
\section{Conclusions}\label{sec:conclusion}
We present finite-blocklength achievability bounds for the \emph{unsourced} A-channel, and we propose easy-to-evaluate refined asymptotic approximations, which are accurate from blocklengths as small as $n=10$. Motivated by the analytical solution of the finite-blocklength bounds and the connection between \gls{ura} and group testing through the unsourced A-channel, we introduce improved decoding algorithms of the so-called tree codes used as part of coding schemes for URA. We show that the proposed decoding algorithms allow to improve the rates achieved by off-the-shelf tree codes significantly at the cost of a moderate increase in decoding complexity. Finally, we adapt our finite-blocklength bounds so that they can be compared against well-known group-testing bounds and schemes. We show that A-channel constructions can perform close to constant tests-per-item constructions, albeit with a much more structured test matrix, which can enable its use in applications such as big data and computer science that usually demand stringent recovery times. For example, A-channel tree-codes test-matrices can be constructed in $\mathcal{O}(\log n)$ time, and the defective set can be reconstructed in $\mathcal{O}(d^2 \log n)$ time. 

\appendices
\section{Proofs of Achievability bounds} \label{app:proof_ach}
\subsection{Preliminaries}\label{app:proof_prel}
%
%
In both error definitions~\eqref{eq:pupe_def} and~\eqref{eq:jpe_def}, we assumed that any collision among the transmitted codewords automatically results in error. It follows that
\begin{equation}\label{eq:penalty_M}
    \prob{\cup_{j\neq i} \{W_j = W_i\}} \leq \frac{\binom{K}{2}}{M}.
\end{equation}
We shall replace the measure under which~\eqref{eq:pupe_def} and~\eqref{eq:jpe_def} are computed by the one under which $\{W_j\}_{j=1}^{K}$ are uniformly sampled without replacement from $[M]$, at the expense of adding a penalty term equal to $\binom{K}{2}/M$ to the upper bounds on the error probability. 

Due to symmetry, we assume without loss of generality that the first $K$ codewords are transmitted. 
For any set $S\in[M]$, let $\vecc(S) \triangleq \bigcup_{j\in S}\vecc_j$. Similarly, for any set $S\in[M]$, we shall use $c_i(S)$ to denote $\bigcup_{j\in S}c_{i,j}$, where $c_{i,j}$ indicates the input of $\vecc_j$ at channel use $i$. We shall omit the subindeces $i$ and $j$ when immaterial. Furthermore, we let $S_\ell$ denote a generic subset of $K-\ell$ elements in $[K]$, and $S'_\ell$ denote a generic subset of $\ell$ elements in $[M]\backslash[K]$.  

Finally, the following definition will turn out useful throughout the proofs. Let $\setA_k \triangleq \{i\in[n]: |Y_i| = k\}$ 
for $k \in [K]$. In words, $\setA_k$ is the set of channel uses where the channel output $\randvecy$ has cardinality $k$. Note that $\sum_{k = 1}^K |\setA_k| = n$. Hence, $A_k$ is the $k$-th element of $\randveca = \tp{[A_1, \dots, A_K]}$, which is a multinomial-distributed random vector with parameters $n$ and $\{p_k\}_{k=1}^K$, where $n$ denotes the number of trials, $K$ the number of possible outcomes in each trial, and $p_k$ the probability that the cardinality of the output is $k$ at channel use $i$, which is given by 
\begin{equation}
    p_k = \prob{|Y_i| = k} = \frac{q!S(K,k)}{q^K(q-k)!}.\label{eq:pk_proof}
\end{equation}
\subsection{Proof of Theorem~\ref{th:ach_A_ch}} \label{app:proof_ach_A_ch}
It follows that
\begin{IEEEeqnarray}{lCl}
    P_{\rm e}^{\rm (p)} &\leq& \Ex{}{\frac{N_{\rm fa,c}}{K+N_{\rm fa,c}}} + \frac{\binom{K}{2}}{M}\label{eq:Pe_nfa_init} \\
    &=& \sum_{\ell = 1}^{M-K} \frac{\ell}{K+\ell} \prob{N\sub{fa,c} = \ell} + \frac{\binom{K}{2}}{M} \label{eq:Pe_nfa_1}\\
    &\leq& \sum_{\ell = 1}^{K-1} \frac{\ell}{K+\ell} \prob{N\sub{fa,c} \geq \ell} + \prob{N\sub{fa,c} \geq K} +  \frac{\binom{K}{2}}{M}. \IEEEeqnarraynumspace\label{eq:Pe_nfa}
\end{IEEEeqnarray}
Hence, to complete the proof of Theorem~\ref{th:ach_A_ch}, we next show that
\begin{IEEEeqnarray}{lCl}
    \prob{N\sub{fa,c}\geq \ell}
    &\leq&
    \mathbb{E}\Biggl[\min\Biggl\{1,\binom{M-K}{\ell}\prod_{k=1}^{K}\lro{\frac{k}{q}}^{\ell A_k}\Biggr\}\Biggr].\IEEEeqnarraynumspace
\end{IEEEeqnarray}
Since the messages are independent and uniform on $[M]$ (see Def.~\ref{def:code}), it follows that
\begin{IEEEeqnarray}{lCl}
    \prob{N\sub{fa,c}\geq \ell} &=&\prob{\bigcup_{S'_\ell} \vecc(S'_\ell) \in \randvecy}\label{eq:more1error}\\
    &=& \prob{\bigcup_{S'_\ell}\bigcap_{k\in[K]}\bigcap_{i\in\setA_k}\lrbo{c_{i}(S'_\ell)\in Y_i}}.\IEEEeqnarraynumspace
\end{IEEEeqnarray}
We next use that $\{\setA_k\}_{k=1}^K$ are disjoint sets together with the law of total probability to write
\begin{IEEEeqnarray}{lCl}
\IEEEeqnarraymulticol{3}{l}{
    \prob{N\sub{fa,c}\geq \ell}}\nonumber\\\quad
    &=& \Ex{\randveca}{  \prob{\bigcup_{S'_\ell}\bigcap_{k\in[K]}\bigcap_{i\in\setA_k}\{c_{i}(S'_\ell)\in Y_i\} \bigm| |\setA_k|=A_k}}\label{eq:E_A_genKq_int}\IEEEeqnarraynumspace\\
    &\leq&\Ex{\randveca}{  \min\lrbo{1, \binom{M-K}{\ell}\prod_{k=1}^{K}\lro{\prob{c(S'_\ell)\in Y}}^{A_k}}}  \label{eq:bound_for_each_list_type} \\
    &=&  \Ex{\randveca}{\min\lrbo{1,\binom{M-K}{\ell}\prod_{k=1}^{K}\lro{\frac{k}{q}}^{\ell A_k}}}\label{eq:E_A_genKq}
\end{IEEEeqnarray}
where the first inequality follows from the union bound, because the messages are independent and uniform on $[M]$ (see Def.~\ref{def:code}), and because the probability that $c_{i}(S'_\ell)\in Y_i$ is independent of $i$, which also justifies why we omitted the subscript $i$. Finally,~\eqref{eq:E_A_genKq} follows since
\begin{IEEEeqnarray}{lCl}
    \prob{ c(S'_\ell) \in Y} &=& \prob{\bigcap_{j\in S'_\ell} \{c_{j} \in Y\}} = (\prob{\bar{c} \in Y})^\ell \IEEEeqnarraynumspace
\end{IEEEeqnarray}
for some generic non-transmitted symbol $\bar{c}$, and because $\prob{ \bar{c} \in Y} = k/q $.
%
\subsection{Proof of Theorem~\ref{th:ach_A_ch_joint}} \label{app:proof_ach_A_ch_joint}
It follows that
\begin{IEEEeqnarray}{lCl}
    \prob{N\sub{fa,j}= \ell} &\leq&\prob{\bigcup_{S_\ell} \bigcup_{S'_\ell} \lrbo{ \vecc(S_\ell) \cup \vecc(S'_\ell) = \randvecy}} + \frac{\binom{K}{2}}{M}\label{eq:more1error_joint}\IEEEeqnarraynumspace
\end{IEEEeqnarray}
and
\begin{IEEEeqnarray}{lCl}
\IEEEeqnarraymulticol{3}{l}{
    \prob{\bigcup_{S_\ell} \bigcup_{S'_\ell}  \lrbo{ \vecc(S_\ell) \cup \vecc(S'_\ell) = \randvecy}}}\nonumber\\\,
    &=& \mathbb{E}_{\randveca}\Biggl[\mathbb{P}\Biggl[\bigcup_{S_\ell} \bigcup_{S'_\ell} \bigcap_{k=1}^{K}\bigcap_{i\in\setA_k}\nonumber\\
    &&\qquad\qquad\quad {} \lrbo{ c_i(S_\ell) \cup c_i(S'_\ell) = \randvecy} \Bigm| |\setA_k|=A_k\Biggr]\Biggr].\IEEEeqnarraynumspace\label{eq:conditional_joint}
\end{IEEEeqnarray}
Since the messages are independent and uniform on $[M]$ (see Def.~\ref{def:code}), by applying the union bound on the right-hand side of~\eqref{eq:conditional_joint}, we have
\begin{IEEEeqnarray}{lCl}
\IEEEeqnarraymulticol{3}{l}{
    \prob{\bigcup_{S_\ell} \bigcup_{S'_\ell}  \lrbo{ \vecc(S_\ell) \cup \vecc(S'_\ell) = \randvecy}}}\nonumber\\\quad
    &\leq& \mathbb{E}\Biggl[\min\Biggl\{1, \binom{K}{K-\ell} \binom{M-K}{\ell} \nonumber\\
    &&\times \prob{\bigcap_{k=1}^{K}\bigcap_{i\in\setA_k}\lrbo{ c_i(S_\ell) \cup c_i(S'_\ell) = Y_i}\Bigm| |\setA_k|=A_k}\Biggr\}\Biggr]\nonumber\\
    \\
    &=& \mathbb{E}\Biggl[\min\Biggl\{1, \binom{K}{K-\ell} \binom{M-K}{\ell}\nonumber\\
    &&\qquad\qquad\times \prod_{k=1}^{K}\lro{\prob{\lrbo{ c(S_\ell) \cup c(S'_\ell) = Y}}}^{A_k}\Biggr\}\Biggr]\label{eq:inter1_genKq_joint}
\end{IEEEeqnarray}
where the last equality follows because $\{\setA_k\}_{k=1}^K$ are disjoint sets together with the law of total probability, and because the considered input distribution is a product distribution. We conclude the proof by showing that
\begin{IEEEeqnarray}{lCl}
    \prob{\lrbo{ c(S_\ell) \cup c(S'_\ell) = Y}} = \sum\limits_{\eta=\underline{\eta}}^{\overline{\eta}}\bar{p}_\eta p(k,\ell,\eta).\IEEEeqnarraynumspace
\end{IEEEeqnarray}

Recall that, in the statement of Theorem~\ref{th:ach_A_ch_joint}, we defined $\eta = |c(S_\ell)|$, i.e., the cardinality of the subset of transmitted codewords $c(S_\ell)$ at a given channel use. Furthermore, we defined $\underline{\eta} = \max\{0,k-\ell\}$ and $\overline{\eta} = \min\{k,K-\ell\}$, where $\ell\in[K]$ denotes the number of elements from the subset of non-transmitted codewords $c(S'_\ell)$. Thus, $K-\ell$ corresponds to the number of elements from the subset of transmitted codewords $c(S_\ell)$. In words, $\underline{\eta}$ represents the minimum number of symbols in channel uses of cardinality $k$ that that need to be covered by the subset of $K-\ell$ transmitted symbols to create a valid output together with the symbols of the subset of $\ell$ non-transmitted codewords. Similarly, $\overline{\eta}$ represents the maximum number of symbols that the subset of $K-\ell$ transmitted codewords could hit in channel uses of cardinality $k$, when we consider $\ell$ non-transmitted codewords. 
The probability term in~\eqref{eq:inter1_genKq_joint} can be expressed as
\begin{IEEEeqnarray}{lCl}
\IEEEeqnarraymulticol{3}{l}{
\prob{\lrbo{ c(S_\ell) \cup c(S'_\ell)} = Y}}\nonumber\\\qquad 
&=& \sum\limits_{\eta=\underline{\eta}}^{\overline{\eta}}\prob{\lrbo{ c(S_\ell) \cup c(S'_\ell)} = Y\bigm| |c(S_\ell)|=\eta}\nonumber\\
    &&\qquad\quad \times \prob{|c(S_\ell)|=\eta}\label{eq:cond_eta_joint}\IEEEeqnarraynumspace\\
    &=&\sum\limits_{\eta=\underline{\eta}}^{\overline{\eta}}\bar{p}_\eta p(k,\ell,\eta).\label{eq:solutio_prob_joint}
\end{IEEEeqnarray}
where $p(k,\ell,\eta) = \prob{\lrbo{ c(S_\ell) \cup c(S'_\ell) = Y}\bigm| |c(S_\ell)|=\eta}$, and
\begin{equation}
    \bar{p}_\eta = \prob{|c(S_\ell)|=\eta} = \frac{k! S(K-\ell,\eta)}{(k-\eta)!  k^{K-\ell} Z_\eta}
\end{equation}
with $Z_\eta$ being a normalizing constant used to make sure that $\sum_{\eta=\underline{\eta}}^{\overline{\eta}} \bar{p}_\eta = 1$. Note that $\bar{p}_\eta$ is similar to $p_k$ in~\eqref{eq:pk_proof}, except that in $\bar{p}_\eta$ not all the values of $\eta\in[K-\ell]$ are possible, since we are considering channel uses of cardinality $k$, and the number of symbols hit by the subset of transmitted codewords needs to be sufficiently large so that the subset of $\ell$ non-transmitted codewords can hit the remaining symbols. Also, $\eta$ cannot be larger than the cardinality $k$. This implies that without $Z_\eta$, $\sum_{\eta=\underline{\eta}}^{\overline{\eta}} \bar{p}_\eta$ could be different from one. 

Given the definition of $p(k,\ell,\eta)$ in~\eqref{eq:p_coupon}, it remains to show how to compute $\pi(k,\ell,\eta)$, which can be computed in closed form only when $\eta=0$ (see Theorem~\ref{th:ach_A_ch_joint}). In Appendix~\ref{app:coupon_gen_eta}, we present a possible way to compute $\pi(k,\ell,\eta)$ for $\eta>0$.   
\section{Computation of $\pi(k,\ell,\eta)$}\label{app:coupon_gen_eta}
Recall that $\pi(k,\ell,\eta)$ denotes the conditional probability, given that $c(S'_\ell)\in Y$, that the subset of non-transmitted symbols 
cover the remaining $k-\eta$ symbols. This problem resembles the classical coupon collector problem where $\pi(k,\ell,\eta)$
is exactly the probability to draw $k$ out of $k$ coupons in $\ell$ steps when one starts with $\eta$ coupons and each coupon appears with
probability $1/k$. 
The evolution of coupons can be modeled by the Markov Chain depicted in Fig.~\ref{fig:markov_chain}. The inter-arrival times
in this chain are independent geometrically distributed random variables. The probability generating function
of the final arrival time can be expressed as
\begin{equation}
    G_{k,\ell,\eta}(z) = z^{k-\eta}\frac{(k-\eta)!}{k^{k-\eta}}\prod_{i=\eta}^k\frac{1}{1-\frac{i}{k}z}.
\end{equation}
Finally, $\pi(k,\ell,\eta)$ can be obtained as the sum of the first $\ell$ coefficients of the polynomial representation of $G_{k,\ell,\eta}(z)$.
These terms can be calculated recursively to avoid numerical issues.
Since $G_{k,\ell,k}(z) = 1$ we have $\pi(k,\ell,k) = 1$. Then
\begin{IEEEeqnarray}{lCl}
    G_{k,\ell,\eta-1}(z) &= z\frac{k-\eta}{k}\left(1-\frac{\eta}{k}z\right)^{-1}G_{k,\ell,\eta}(z)\\
     &= z\frac{k-\eta}{k}\sum_{i=0}^\infty \left(\frac{\eta}{k}z\right)^iG_{k,\ell,\eta}(z).
\end{IEEEeqnarray}
Therefore, the polynomial representation of $G_{k,\ell,\eta-1}(z)$ can be computed
from $G_{k,\ell,\eta}(z)$ by convolution with the polynomial $\sum_{i=0}^\infty \left(\frac{\eta}{k}z\right)^i$. Note that only the first $\ell$ coefficients of $G_{k,\ell,\eta}(z)$ are relevant, so it suffices to compute the convolution
with $\sum_{i=0}^\ell \left(\frac{\eta}{k}z\right)^i$.
%
\section{Proof of \eqref{eq:MI_ineq}}
\label{sec:MI_ineq_proof}
By symmetry we write $\mu_\ell = I(\randvecx_{[\ell]};Y|\randvecx_{[\ell+1:K]})$. We next show that
\begin{equation}
	\frac{\mu_\ell}{\ell} \geq \frac{\mu_K}{K}
\end{equation}
for every $\ell\leq K$, which implies \eqref{eq:MI_ineq}.

First, note that, since all $X_i$ are iid, it holds that $I(\randvecx_S;Y|\randvecx_{S'}) \leq I(\randvecx_S;Y|\randvecx_{S''})$ for $S,S’,S'' \subset [K]$ whenever $S\cap S' = S\cap S'' = \emptyset$ and $S’\subset S''$ which follows from $I(X_2;Y)\leq I(X_2;Y|X_1)$ for independent $X_1,X_2$. In other words, for independent random variables, conditioning increases mutual information. The latter follows from the convexity of $I(X_2;Y)$ in $p(y|x_2) = \sum_{x_1} p(y|x_1,x_2)p(x_1)$. Second, again, due to the iid property, the elements of $\randvecx$ can be arbitrary permuted. With these two properties and repeated use of the chain rule for mutual information we can show that $\frac{\mu_\ell}{\ell} \geq \frac{\mu_{\ell+1}}{\ell+1}$:
	\begin{equation}
		\begin{split}
			\ell\mu_{\ell+1} &= \ell I(\randvecx_{[\ell]};Y|\randvecx_{[\ell+1:K]}) + \ell I(X_{\ell+1};Y|\randvecx_{[\ell+2:K]})\\
			 	&= \ell\mu_\ell + \ell I(X_{1};Y|\randvecx_{[\ell+2:K]}).
		\end{split}
		\label{eq:Il}
	\end{equation}
	By the chain rule $\mu_\ell$ can be expressed as
	\begin{equation}
		\mu_\ell = \sum_{i=1}^\ell I(X_i;Y|\randvecx_{[i+1:K]}).
	\end{equation}
	It is apparent that the righ-hand side of \eqref{eq:Il} can be upper bound by $\mu_\ell$ by conditioning on additional $X_i$’s, which shows that $\ell\mu_{\ell+1} \leq (\ell+1)\mu_\ell$.

%
\section*{Acknowledgement}

The authors gratefully acknowledge fruitful discussions with Khac-Hoang Ngo.

\bibliographystyle{IEEEtran}

\begin{thebibliography}{}
\providecommand{\url}[1]{#1}
\csname url@samestyle\endcsname
\providecommand{\newblock}{\relax}
\providecommand{\bibinfo}[2]{#2}
\providecommand{\BIBentrySTDinterwordspacing}{\spaceskip=0pt\relax}
\providecommand{\BIBentryALTinterwordstretchfactor}{4}
\providecommand{\BIBentryALTinterwordspacing}{\spaceskip=\fontdimen2\font plus
\BIBentryALTinterwordstretchfactor\fontdimen3\font minus
  \fontdimen4\font\relax}
\providecommand{\BIBforeignlanguage}[2]{{%
\expandafter\ifx\csname l@#1\endcsname\relax
\typeout{** WARNING: IEEEtran.bst: No hyphenation pattern has been}%
\typeout{** loaded for the language `#1'. Using the pattern for}%
\typeout{** the default language instead.}%
\else
\language=\csname l@#1\endcsname
\fi
#2}}
\providecommand{\BIBdecl}{\relax}
\BIBdecl

\end{thebibliography}


\begin{thebibliography}{10}
    \providecommand{\url}[1]{#1}
    \csname url@samestyle\endcsname
    \providecommand{\newblock}{\relax}
    \providecommand{\bibinfo}[2]{#2}
    \providecommand{\BIBentrySTDinterwordspacing}{\spaceskip=0pt\relax}
    \providecommand{\BIBentryALTinterwordstretchfactor}{4}
    \providecommand{\BIBentryALTinterwordspacing}{\spaceskip=\fontdimen2\font plus
    \BIBentryALTinterwordstretchfactor\fontdimen3\font minus
      \fontdimen4\font\relax}
    \providecommand{\BIBforeignlanguage}[2]{{%
    \expandafter\ifx\csname l@#1\endcsname\relax
    \typeout{** WARNING: IEEEtran.bst: No hyphenation pattern has been}%
    \typeout{** loaded for the language `#1'. Using the pattern for}%
    \typeout{** the default language instead.}%
    \else
    \language=\csname l@#1\endcsname
    \fi
    #2}}
    \providecommand{\BIBdecl}{\relax}
    \BIBdecl
    
    \bibitem{Chang81}
    S.-C. Chang and J.~Wolf, ``On the {T}-user {M}-frequency noiseless
      multiple-access channel with and without intensity information,''
      \emph{{IEEE} Trans. Inf. Theory}, vol.~27, no.~1, pp. 41--48, Jan. 1981.
    
    \bibitem{Bassalygo2000}
    \BIBentryALTinterwordspacing
    L.~Bassalygo and M.~Pinsker, \emph{Calculation of the Asymptotically Optimal
      Capacity of a T-User M-Frequency Noiseless Multiple-Access Channel}.\hskip
      1em plus 0.5em minus 0.4em\relax Boston, MA: Springer US, 2000, pp. 177--180.
      [Online]. Available: \url{https://doi.org/10.1007/978-1-4757-6048-4_16}
    \BIBentrySTDinterwordspacing
    
    \bibitem{Bas2013}
    L.~A. Bassalygo and V.~V. Rykov, ``Multiple-access hyperchannel,''
      \emph{Problems of Information Transmission}, vol.~49, no.~4, pp. 299--307,
      Oct. 2013.
    
    \bibitem{Fengler21-05}
    A.~Fengler, P.~Jung, and G.~Caire, ``{SPARC}s for unsourced random access,''
      \emph{{IEEE} Trans. Inf. Theory}, vol.~67, no.~10, pp. 6894--6915, May 2021.
    
    \bibitem{polyanskiy17-06a}
    Y.~Polyanskiy, ``A perspective on massive random-access,'' in \emph{Proc. IEEE
      Int. Symp. Inf. Theory (ISIT)}, Jun. 2017, pp. 2523--2527.
    
    \bibitem{Zadik19}
    I.~Zadik, Y.~Polyanskiy, and C.~Thrampoulidis, ``Improved bounds on {G}aussian
      {MAC} and sparse regression via {G}aussian inequalities,'' in \emph{Proc.
      IEEE Int. Symp. Inf. Theory (ISIT)}, Jul. 2019, pp. 430--434.
    
    \bibitem{Kowshik21}
    S.~S. Kowshik and Y.~Polyanskiy, ``Fundamental limits of many-user {MAC} with
      finite payloads and fading,'' \emph{{IEEE} Trans. Inf. Theory}, vol.~67,
      no.~9, pp. 5853--5884, Jun. 2021.
    
    \bibitem{Ngo22}
    \BIBentryALTinterwordspacing
    K.-H. Ngo, A.~Lancho, G.~Durisi, and A.~Graell~i Amat, ``Unsourced multiple
      access with random user activity,'' Feb. 2022. [Online]. Available:
      \url{https://arxiv.org/abs/2202.06365}
    \BIBentrySTDinterwordspacing
    
    \bibitem{Ravi22}
    J.~Ravi and T.~Koch, ``Scaling laws for {G}aussian random many-access
      channels,'' \emph{{IEEE} Trans. Inf. Theory}, vol.~68, no.~4, pp. 2429--2459,
      Apr. 2022.
    
    \bibitem{Pra2021a}
    A.~K. Pradhan, V.~K. Amalladinne, K.~R. Narayanan, and J.-F. Chamberland,
      ``{LDPC} codes with soft interference cancellation for uncoordinated
      unsourced multiple access,'' in \emph{Proc. IEEE Int. Conf. Commun. (ICC)},
      Jun. 2021.
    
    \bibitem{Ama2022}
    V.~K. Amalladinne, A.~K. Pradhan, C.~Rush, J.-F. Chamberland, and K.~R.
      Narayanan, ``Unsourced random access with coded compressed sensing:
      {{Integrating AMP}} and belief propagation,'' \emph{{IEEE} Trans. Inf.
      Theory}, vol.~68, no.~4, pp. 2384--2409, Apr. 2022.
    
    \bibitem{Tru2021a}
    D.~Truhachev, M.~Bashir, A.~Karami, and E.~Nassaji, ``Low-complexity coding and
      spreading for unsourced random access,'' \emph{{IEEE} Commun. Lett.},
      vol.~25, no.~3, pp. 774--778, Mar. 2021.
    
    \bibitem{Ama2020a}
    V.~K. Amalladinne, J.-F. Chamberland, and K.~R. Narayanan, ``A coded compressed
      sensing scheme for unsourced multiple access,'' \emph{{IEEE} Trans. Inf.
      Theory}, vol.~66, no.~10, pp. 6509--6533, Jul. 2020.
    
    \bibitem{Fen2021c}
    A.~Fengler, S.~Haghighatshoar, P.~Jung, and G.~Caire, ``Non-{Bayesian} activity
      detection, large-scale fading coefficient estimation, and unsourced random
      access with a massive {MIMO} receiver,'' \emph{{IEEE} Trans. Inf. Theory},
      vol.~67, no.~5, pp. 2925--2951, May 2021.
    
    \bibitem{And2021}
    K.~Andreev, P.~Rybin, and A.~Frolov, ``Reed-{Solomon} coded compressed sensing
      for the unsourced random access,'' in \emph{in Proc. {IEEE} Int. Symp. Wirel.
      Comm. Syst. (ISWCS)}, Sep. 2021.
    
    \bibitem{Lia2021}
    Z.~Liang, J.~Zheng, and J.~Ni, ``Index modulation--aided mixed massive random
      access,'' \emph{Frontiers in Communications and Networks}, vol.~2, 2021.
    
    \bibitem{Che2022}
    J.~Che, Z.~Zhang, Z.~Yang, X.~Chen, C.~Zhong, and D.~W.~K. Ng, ``Unsourced
      random massive access with beam-space tree decoding,'' \emph{{IEEE} J. Sel.
      Areas Commun.}, vol.~40, no.~4, pp. 1146--1161, Apr. 2022.
    
    \bibitem{Dor1943}
    R.~Dorfman, ``The detection of defective members of large populations,''
      \emph{Annals of Mathematical Statistics}, vol.~14, no.~4, pp. 436--440, Dec.
      1943.
    
    \bibitem{Du2006}
    D.-Z. Du and F.~K. Hwang, \emph{Pooling {{Designs}} and {{Nonadaptive Group
      Testing}}: {{Important Tools}} for {{DNA Sequencing}}}, ser. Series on
      {{Applied Mathematics}}.\hskip 1em plus 0.5em minus 0.4em\relax {WORLD
      SCIENTIFIC}, Jun. 2006, vol.~18.
    
    \bibitem{Ama2019a}
    V.~K. Amalladinne, K.~R. Narayanan, J.-F. Chamberland, and D.~Guo,
      ``Asynchronous neighbor discovery using coupled compressive sensing,'' in
      \emph{Proc. IEEE Int. Conf. Acoust., Speech, Signal Process. (ICASSP)}, May
      2019, pp. 4569--4573.
    
    \bibitem{Ber1984a}
    T.~Berger, N.~Mehravari, D.~Towsley, and J.~Wolf, ``Random multiple-access
      communication and group testing,'' \emph{{IEEE} Trans. Commun.}, vol.~32,
      no.~7, pp. 769--779, Jul. 1984.
    
    \bibitem{Mal2013}
    D.~M. Malioutov and K.~R. Varshney, ``Exact rule learning via {Boolean}
      compressed sensing,'' in \emph{Proc. Int. Conf. Machine Learning}, ser.
      {{ICML}}'13, vol.~28.\hskip 1em plus 0.5em minus 0.4em\relax {Atlanta, GA,
      USA}: {JMLR.org}, Jun. 2013, pp. III--765--III--773.
    
    \bibitem{Xua2010}
    Y.~Xuan, I.~Shin, M.~T. Thai, and T.~Znati, ``Detecting application
      denial-of-service attacks: {A} group-testing-based approach,'' \emph{{IEEE}
      Trans. Parallel Distrib. Syst.}, vol.~21, no.~8, pp. 1203--1216, Aug. 2010.
    
    \bibitem{Ald2019a}
    M.~Aldridge, O.~Johnson, and J.~Scarlett, ``Group {{Testing}}: {{An Information
      Theory Perspective}},'' \emph{Foundations and Trends\textregistered{} in
      Communications and Information Theory}, vol.~15, no. 3-4, pp. 196--392, Dec.
      2019.
    
    \bibitem{Joh2019}
    O.~Johnson, M.~Aldridge, and J.~Scarlett, ``Performance of group testing
      algorithms with near-constant tests-per-item,'' \emph{{IEEE} Trans. Inf.
      Theory}, vol.~65, no.~2, pp. 707--723, Feb. 2019.
    
    \bibitem{Kau1964}
    W.~Kautz and R.~Singleton, ``Nonrandom binary superimposed codes,''
      \emph{{IEEE} Trans. Inf. Theory}, vol.~10, no.~4, pp. 363--377, Oct. 1964.
    
    \bibitem{Ina2018}
    H.~A. Inan, P.~Kairouz, M.~Wootters, and A.~Özgür, ``On the optimality of the
      {Kautz-Singleton} construction in probabilistic group testing,'' \emph{{IEEE}
      Trans. Inf. Theory}, vol.~65, no.~9, pp. 5592--5603, Mar. 2019.
    
    \bibitem{olver10}
    F.~W.~J. Olver, D.~W. Lozier, R.~F. Boisvert, and C.~W. Clark, \emph{The {NIST}
      Handbook of Mathematical Functions}.\hskip 1em plus 0.5em minus 0.4em\relax
      Cambridge Univ. Press, 2010.
    
    \bibitem{feller71-a}
    W.~Feller, \emph{An Introduction to Probability Theory and Its Applications},
      2nd~ed.\hskip 1em plus 0.5em minus 0.4em\relax New York, NY, USA: Wiley,
      1971, vol.~II.
    
    \bibitem{polyanskiy10-05a}
    Y.~Polyanskiy, H.~V. Poor, and S.~Verd\'u, ``Channel coding rate in the finite
      blocklength regime,'' \emph{{IEEE} Trans. Inf. Theory}, vol.~56, no.~5, pp.
      2307--2359, May 2010.
    
    \bibitem{Ald2014}
    M.~Aldridge, L.~Baldassini, and O.~Johnson, ``Group testing algorithms:
      {Bounds} and simulations,'' \emph{{IEEE} Trans. Inf. Theory}, vol.~60, no.~6,
      pp. 3671--3687, Jun. 2014.
    
    \bibitem{Mal2012}
    D.~Malioutov and M.~Malyutov, ``Boolean compressed sensing: {{LP}} relaxation
      for group testing,'' in \emph{Proc. IEEE Int. Conf. Acoust., Speech, Signal
      Process. (ICASSP)}, Mar. 2012, pp. 3305--3308.
    
    \bibitem{cover06-a}
    T.~M. Cover and J.~A. Thomas, \emph{Elements of Information Theory},
      2nd~ed.\hskip 1em plus 0.5em minus 0.4em\relax New York, NY, U.S.A.: Wiley,
      2006.
    
    \end{thebibliography}

\end{document}